\documentclass[useAMS,usenatbib]{mnras}
\usepackage{graphicx}
\usepackage{times}
\usepackage{amssymb}
\usepackage{amsmath}
\usepackage{multirow}
\usepackage{subfig}
\usepackage{hyphenat}
\usepackage{hyperref}
\usepackage{sansmath}
\usepackage{aas_macros}
\usepackage{tabularx}
\newcommand{\hompc}{\,h\,{\rm Mpc}^{-1}}
\newcommand{\mpcoh}{\,h^{-1}\,{\rm Mpc}}

\newcommand{\hompcV}{\,h^{3}\,{\rm Mpc^{-3}}}

\newcommand{\bk}{\boldsymbol{k}}

\newcommand{\bx}{\boldsymbol{x}}
\newcommand{\br}{\boldsymbol{r}}
\newcommand{\bv}{\boldsymbol{v}}

\newcommand{\bC}{\boldsymbol{\mathsf{C}}}

\begin{document}

\title[2MTF Velocity Power Spectrum] 
{2MTF VI. Measuring the velocity power spectrum.}

\author[C. Howlett et. al.]{\parbox{\textwidth}{
Cullan Howlett\thanks{Email: cullan.howlett@icrar.org}$^{1,2}$,
Lister Staveley-Smith$^{1,2}$,
Pascal J. Elahi$^{1,2}$,
Tao Hong$^{2,3}$,
Tom H. Jarrett$^{4}$,
D. Heath Jones$^{5}$,
B{\"a}rbel S. Koribalski$^{6}$,
Lucas M. Macri$^{7}$,
Karen L. Masters$^{8}$,
Christopher M. Springob$^{1,2}$.
}
  \vspace*{4pt} \\ 
$^{1}$ International Centre for Radio Astronomy Research, The University of Western Australia, Crawley, WA 6009, Australia. \\
$^{2}$ ARC Centre of Excellence for All-sky Astrophysics (CAASTRO). \\
$^{3}$ National Astronomical Observatories, Chinese Academy of Sciences, 20A Datun Road, Chaoyang District, Beijing 100012, China. \\
$^{4}$ Astronomy Department, University of Cape Town, Private Bag X3. Rondebosch 7701, Republic of South Africa. \\
$^{5}$ English Language and Foundation Studies Centre, University of Newcastle, Callaghan, NSW 2308, Australia. \\
$^{6}$ CSIRO Astronomy \& Space Science, Australia Telescope National Facility, PO Box 76, Epping, NSW 1710, Australia. \\
$^{7}$ George P. and Cynthia Woods Mitchell Institute for Fundamental Physics and Astronomy, Department of Physics and Astronomy, \\
Texas A\&M University, 4242 TAMU, College Station, TX 77843, USA. \\
$^{8}$ Institute of Cosmology \& Gravitation, Dennis Sciama Building, University of Portsmouth, Portsmouth, PO1 3FX, UK.
}

\pagerange{\pageref{firstpage}--\pageref{lastpage}} \pubyear{2016}
\maketitle
\label{firstpage}

\begin{abstract}
We present measurements of the velocity power spectrum and constraints on the growth rate of structure $f\sigma_{8}$, at redshift zero, using the peculiar motions of 2,062 galaxies in the completed 2MASS Tully-Fisher survey (2MTF). To accomplish this we introduce a model for fitting the velocity power spectrum including the effects of non-linear Redshift Space Distortions (RSD), allowing us to recover unbiased fits down to scales $k=0.2\hompc$ without the need to smooth or grid the data. Our fitting methods are validated using a set of simulated 2MTF surveys. Using these simulations we also identify that the Gaussian distributed estimator for peculiar velocities of \cite{Watkins2015} is suitable for measuring the velocity power spectrum, but sub-optimal for the 2MTF data compared to using magnitude fluctuations $\delta m$, and that, whilst our fits are robust to a change in fiducial cosmology, future peculiar velocity surveys with more constraining power may have to marginalise over this. We obtain \textit{scale-dependent} constraints on the growth rate of structure in two bins, finding $f\sigma_{8} = [0.55^{+0.16}_{-0.13},0.40^{+0.16}_{-0.17}]$ in the ranges $k = [0.007-0.055, 0.55-0.150]\hompc$. We also find consistent results using four bins. Assuming scale-\textit{independence} we find a value $f\sigma_{8} = 0.51^{+0.09}_{-0.08}$, a $\sim16\%$ measurement of the growth rate. Performing a consistency check of General Relativity (GR) and combining our results with CMB data only we find $\gamma = 0.45^{+0.10}_{-0.11}$, a remarkable constraint considering the small number of galaxies. All of our results are completely independent of the effects of galaxy bias, and fully consistent with the predictions of GR (scale-independent $f\sigma_{8}$ and $\gamma\approx0.55$).
\end{abstract}


\section{Introduction}
The current cosmological paradigm consists of a flat, $\Lambda$CDM universe, whose geometry and structure evolved according to the equations of General Relativity (GR, \citealt{Einstein1916}) and a matter-radiation tensor dominated by dark energy and cold dark matter. Recent observations of the Cosmic Microwave Background radiation (CMB; \citealt{Planck2016}), Large Scale Structure (e.g., \citealt{Alam2016}), Type Ia supernovae (e.g., \citealt{Riess2016}) and weak gravitational lensing (\citealt{Heymans2012}), provide strong support for this cosmological model and confirm the presence of both dark energy and dark matter. 
However, the fact remains that there is currently no convincing physical explanation for the nature of the ``dark universe''. This indicates that our current understanding of particle physics or gravitation is lacking, and further tests or observations are required to determine how.
 
One such test involves studying the relationship between density and velocity in our universe. Assuming GR, the velocity with which an object at position $\bx$ and scale factor $a$ moves, $\bv(\bx,a)$, is related to the velocity divergence field, $\theta(\bx,a)$, via the continuity equation,
\begin{equation}
\nabla \cdot \bv(\bx,a) = aH(a)f(a)\theta(\bx,a).
\label{eq:continuity}
\end{equation}
On linear scales the velocity divergence field is equal (but opposite in sign) to the density field $\delta(\bx,a)$. The key parameters associated with this equation are the Hubble parameter, $H(a)$, which depends on the expansion rate today and the relative densities of dark energy and matter, and the linear growth rate $f=\mathrm{d\,ln\,}D/\mathrm{d\,ln\,}a$, which is the change in the linear growth factor over time. The $\Lambda$CDM+GR model provides strong predictions for these. In particular GR predicts that the growth rate is scale-independent and evolves with time as $f=\Omega_{m}(a)^{\gamma}$, where $\gamma \approx 0.55$ and $\Omega_{m}(a)$ is the matter density of the Universe \citep{Linder2007}. If the laws governing the relation between velocity and density in our Universe differ from GR, one could expect to find some deviation in the form of $f$.

Measurements of the density and/or velocity fields can be made using galaxy redshifts. Inhomogeneities in the density field generate gravitational potential wells and induce `peculiar' velocities (PVs) in the galaxies around them. Hence the redshift measured by an observer $z$, is a function of that caused by the expansion of the universe $z_{H}$, and the PV of the observed galaxy along the line-of-sight $v_{pec}$,\footnote{Note that this is only true in the reference frame comoving with the observer (usually called the CMB frame, where the observer's own peculiar velocity is zero), only in the case where we neglect the effects of gravitational lensing caused by the same inhomogeneities that generate the peculiar motions, \textit{and} only if the peculiar velocities are non-relativistic. For a detailed discussion of how these effects change the measured redshift see, for example, \cite{DavisT2011}, \cite{DavisT2014} and \cite{Wojtak2015}.}
\begin{equation}
1+z = (1+z_{H})(1+v_{pec}/c).
\label{eq:redshift}
\end{equation}

Because of Eq.~\ref{eq:redshift}, the peculiar motions of galaxies change the observed clustering of galaxies compared to what would be measured if their true distance were known. This effect is known as Redshift Space Distortion (RSD; \citealt{Kaiser1987}), and can be exploited to obtain constraints on the growth rate. However, galaxies are biased tracers of the matter density \citep{Cole1989,Fry1993} and on linear scales galaxy bias is exactly degenerate with the growth rate. This degeneracy is partially broken on non-linear scales so the efficacy of RSD is limited by how far into the non-linear regime the clustering can be successfully modelled, or how well we know the galaxy bias. Even with this limitation, RSD in the clustering of galaxies has been widely used to measure the growth rate and provides some of the strongest large scale tests of GR to date (see e.g. \citealt{Blake2011,Beutler2012,delaTorre2013,Howlett2015a,Alam2016}).

Alternatively, if the distance to a galaxy is known then we can evaluate $z_{H}$, compare this to its measured redshift to calculate the peculiar velocity and then use this to constrain the growth rate of structure. This can be done by directly comparing the measured linear velocity and density fields \protect{\citep{DavisM1996,Branchini1999,Erdogdu2006,DavisM2011,Branchini2012,Springob2014,Carrick2015,Springob2016}} or by looking at a distribution of peculiar velocities and constraining their 2-point functions \protect{\citep{Gorski1989,Jaffe1995,Silberman2001,Macaulay2012,Johnson2014}}. Methods of determining the true distance to a galaxy include the Tully-Fisher relation (TF; \citealt{Tully1977}), the Fundamental Plane of galaxies \citep{Dressler1987,Djorgovski1987} and the use of supernovae as standard candles \citep{Phillips1993}.

There are several benefits to looking at the velocity power spectrum as opposed to measuring only the redshift-space galaxy clustering or directly comparing the velocity and density fields. Firstly, galaxy velocities are expected to trace the underlying velocity field exactly on large scales \protect{\citep{Tinker2006,Desjacques2010,Elia2012,Jennings2015}}. That is, we can obtain measurements of the growth rate unsullied by the complicated way in which galaxies populate the underlying matter field. Secondly, because the measurement of the growth rate is largely independent of galaxy bias, the velocity power spectrum can be measured on distinct scales, providing constraints on the scale-dependence of the growth rate. As the velocity field probes much larger scales than the density field (which can be seen from the Fourier transform of Eq.~\ref{eq:continuity}), it can also provide constraints on the growth rate of structure outside of the survey window.

In this paper we use the distribution of peculiar velocities in the completed 2MASS Tully-Fisher survey (2MTF; \citealt{Masters2008,Hong2013,Masters2014,Hong2014,Springob2016}, Hong et al., in preparation) to measure the velocity power spectrum and constrain the growth rate of structure. We base our method on the work of \cite{Macaulay2012} and \cite{Johnson2014}, using more realistic simulations to precisely test the robustness of the fits, and add improvements to the non-linear accuracy of the modelling. Compared to other modern, larger, peculiar velocity surveys such as SFI++ \citep{Springob2007} and 6dFGSv \citep{Springob2014}, 2MTF benefits from a near homogeneous full-sky coverage, better fractional distance error, a higher number density of nearby objects where the distance measurements are typically more accurate, and by default has measurements (as does 6dFGSv) presented as `log-distance' ratios rather than velocities, which preserves the Gaussian nature of the measurement errors. Additionally, larger compilations of peculiar velocities, such as CosmicFlows-3 \citep{Tully2016}, could have systematic errors resulting from incorrect calibration of the relative zero-points between the different sub-surveys which would be difficult to test using mocks due to the large number of selection functions that would have to be incorporated into the simulations. 

The layout of this paper is as follows. In Section \ref{sec:data} we present the data we will use and a set of realistic mock surveys generated to test the methods we will apply to the data. In Section \ref{sec:model} we detail the theoretical model for the velocity power spectrum and how the distribution of peculiar velocities can be used to measure this. We apply the method to the mock surveys in Section \ref{sec:tests}, highlightling the regime in which the model works, before applying this same model to the data in Section \ref{sec:results}. We will discuss and compare our results to the predictions from GR and those from other surveys and methods in Section \ref{sec:conclusions}.

Within this paper we adopt a flat, neutrinoless cosmological model, based on the results of \cite{Planck2016}: $\Omega_{m}=0.3121$, $\Omega_{b}=0.0488$, $H_{0}=67.51\mathrm{km\,s^{-1}\,Mpc^{-1}}$, $n_{s}=0.9653$ and $\sigma_{8}=0.815$. For this model, assuming GR, the \textit{expected} value of the normalised growth rate is $f\sigma_{8}=0.432$. Note, however, that in our measurements we do not make such an assumption. We simply adopt a cosmological model, make an independent measurement of the growth rate, and then compare against the GR prediction.

\section{Data and simulations} \label{sec:data}
\subsection{The 2MASS Tully-Fisher Survey}
The completed 2MTF survey (see \citealt{Hong2014,Springob2016} for work using earlier versions of the data and Hong et al., in preparation, for the complete data set) is a survey of 2062 nearby, bright, spiral galaxies with measured redshifts and distances derived using the TF relation. Targets for the 2MTF survey were selected from the 2MASS Redshift Survey (2MRS; \citealt{Huchra2012}) with a limiting total $K$-band magnitude of 11.25, co-added axis ratio $b/a<0.5$ and redshift $cz < 10,000\,\mathrm{km\,s^{-1}}$. The photometric properties for the sample, namely the $J$, $H$ and $K$-band total magnitudes, co-added axis ratios and morphological classifications are taken from the 2MASS Extended Source catalogue \citep{Jarrett2000}. For the $\sim6,000$ galaxies satisfying these selection criteria, HI measurements are obtained for the brightest using a combination of archival data, mainly from the Cornell HI digital archive \citep{Springob2005} but also from other sources; data from the Arecibo Legacy Fast ALFA survey (ALFALFA; \citealt{Giovanelli2005}); and targeted observations using both the Robert C. Byrd Green Bank Telescope (GBT; \citealt{Masters2014}) and Parkes telescope \citep{Hong2013}.

\begin{figure*}
\centering
\includegraphics[width=\textwidth, trim={0 3cm 0 0}]{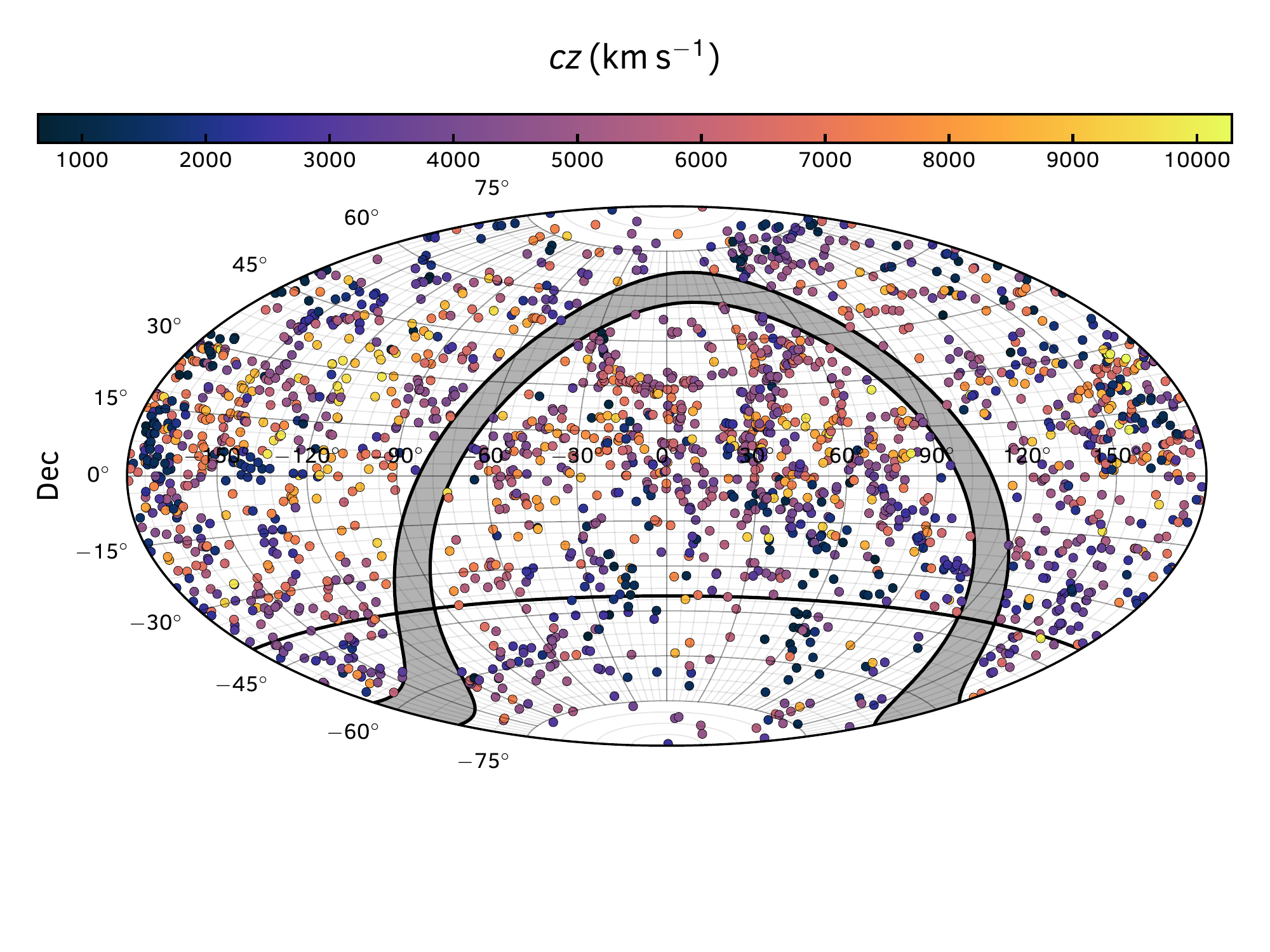} \\
  \caption{The sky coverage of the 2MTF survey in equatorial coordinates. The color of the points represents the redshift of each galaxy. The 2MTF survey is effectively full-sky except for the area within $5^{\circ}$ of the galactic plane (grey shaded area). The dividing line at $\delta=-40^{\circ}$ indicates the region below which the number density of the 2MTF survey falls by a factor of $\sim 2$ due to the different telescopes used to collect the data (c.f. Fig.~\ref{fig:nz}).}
  \label{fig:sky}
\end{figure*}

\begin{figure}
\centering
\includegraphics[width=0.5\textwidth]{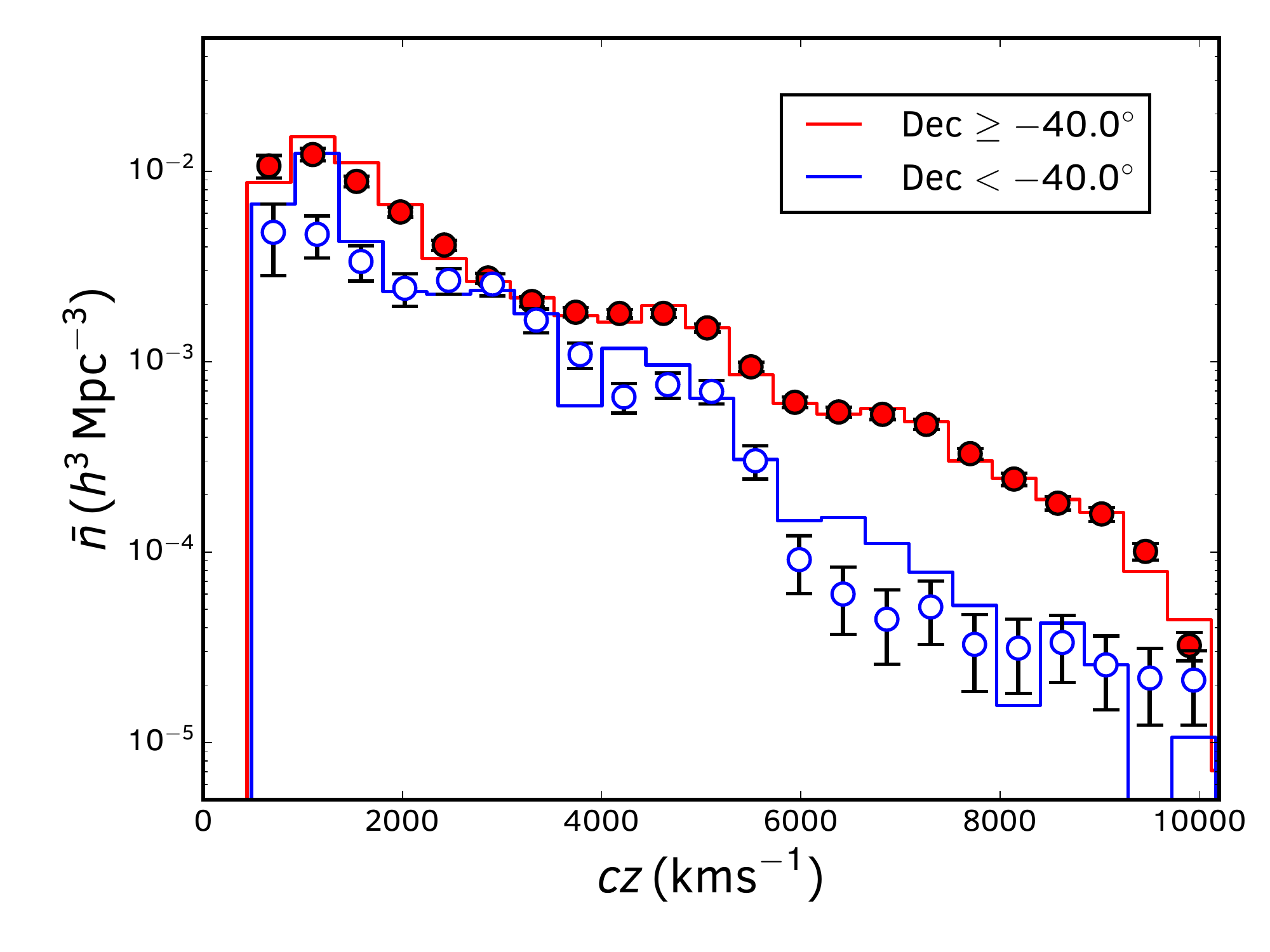} \\
  \caption{The number density of galaxies in the 2MTF dataset (solid lines) and the average and variance of the mock 2MTF catalogues (points, detailed in Section~\ref{sec:mocks}). Red lines and points show the distribution for galaxies north of a declination of $-40.0^{\circ}$, whilst blue shows those south of this. This difference occurs due to the different instruments and telescopes used to obtain the HI line-widths, and the mocks are generated by subsampling these two regions separately.}
  \label{fig:nz}
\end{figure}

Additional cuts are applied to those galaxies with HI measurements to improve the data quality. Only galaxies with $cz > 600\,\mathrm{km\,s^{-1}}$, relative HI line width error less than $10\%$ and HI spectrum signal to noise ratio of $SNR>5$ were included in the final 2MTF sample. The sky coverage and redshift distribution of the 2062 galaxies in the final sample are shown in Fig.~\ref{fig:sky} and \ref{fig:nz}respectively. Due to the different telescopes used to obtain the HI measurements, in particular the use of only the Parkes telescope to obtain measurements for galaxies with declination $\delta<-40.0^{\circ}$, the number density of targets is different above and below this declination. We do not correct for this in our measurements, but our results are shown to be robust to the selection effects within the survey (see Section~\ref{sec:tests}).

Distance measurements to the 2MTF galaxies are obtained by comparing the absolute magnitudes in the $J$, $H$ and $K$-bands to the absolute magnitudes inferred from the TF relation. In the 2MTF catalogue, distances are presented as log-distance ratios, $\Delta d$, the logarithm of the ratio of the distance calculated using the measured redshift, $D_{z}$, and the true comoving distance, $D_{H}$. The Tully-Fisher relation was fit separately for these three bands to a distinct sample of 888 cluster galaxies using a revised version of the method in \cite{Masters2008}. The method for calculating the absolute magnitude of each galaxy, including internal dust and $k$-corrections is also detailed in \cite{Masters2008}.  The log-distance ratio can be calculated from the difference in absolute magnitudes $\Delta M = M_{obs}-M(W)$, where $M_{obs}$ is the observed corrected absolute magnitude and $M(W)$ is that inferred from the TF relation, via
\begin{equation}
\Delta d = \mathrm{log}\left(\frac{D_{z}}{D_{H}}\right) = -\frac{\Delta M}{5}.
\label{eq:mag}
\end{equation}
Hence if the errors in the distance measurements are log-normal, the errors in the log-distance ratio are Gaussian. However, the exact conversion from log-distance ratios to a peculiar velocity is generally non-linear and gives a non-Gaussian PDF for the peculiar velocity (c.f., \citealt{Johnson2014,Springob2014,Scrimgeour2016}). We demonstrate that there are estimators and variable transformations that allow the Gaussian nature of the measurements to be retained and related to the peculiar velocity (see Section~\ref{sec:model}).

\begin{figure*}
\centering
\includegraphics[width=\textwidth]{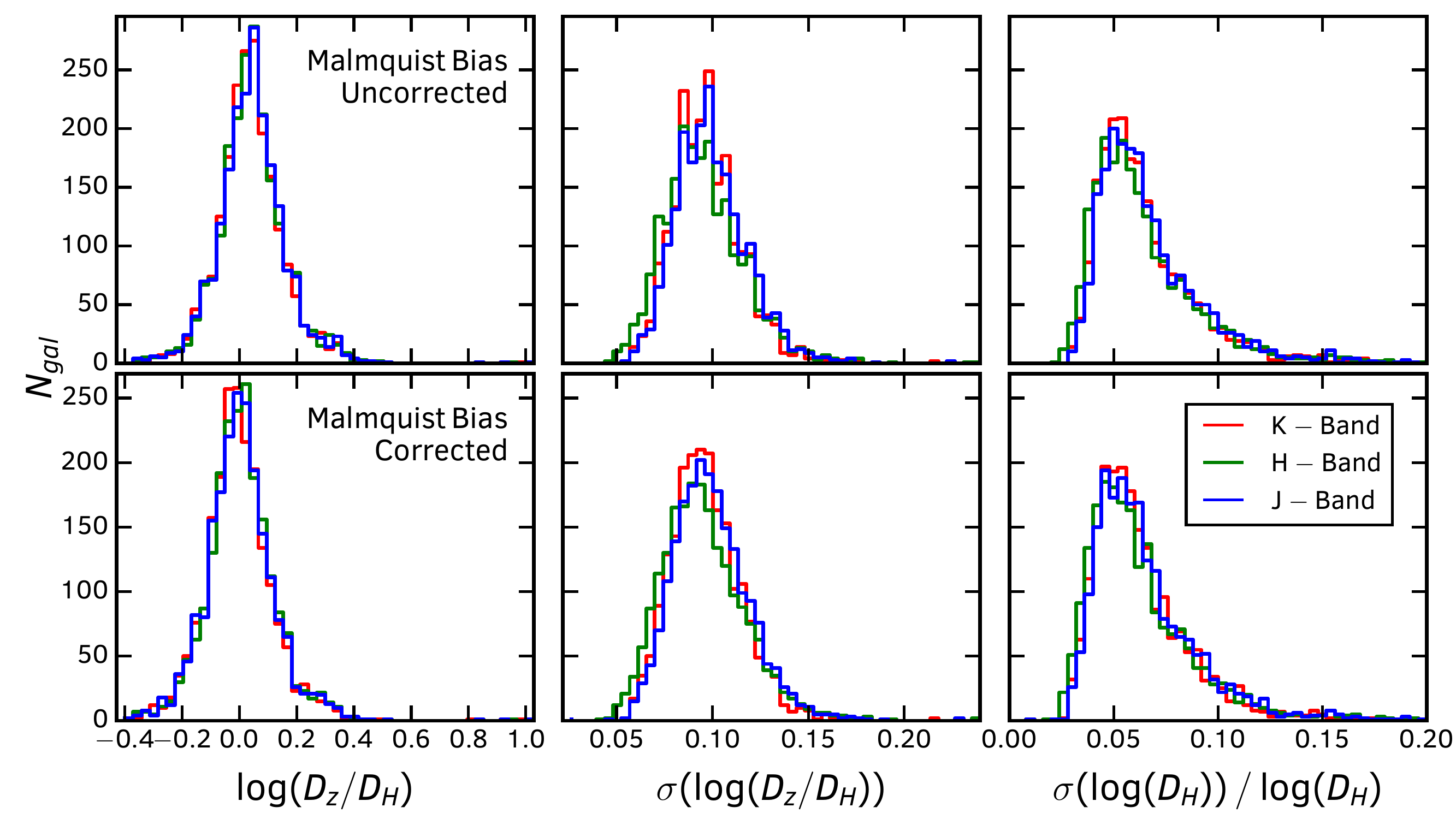}
  \caption{Histograms of the log-distance ratios (left), errors in the log-distance ratios (middle) and relative error in the distances (right) for the 2MTF galaxies. The upper row shows the data before the correction for Malmquist bias, the bottom row is the data after the correction was applied. Different colors show the data from the three separate fits to the TF relation for the three photometric ($J$, $H$ and $K$) bands.
  }
  \label{fig:logdist}
\end{figure*}

Finally, the distance measurements are corrected for homogeneous Malmquist bias \citep{Malmquist1924}, arising from the fact that objects at higher redshift probe larger cosmological volumes, and are often intrinsically brighter than their nearby counterparts to be measurable. This is done using the method described in \cite{Springob2016}. Histograms of the log-distance ratios and their errors in the $J$, $H$ and $K$-bands before and after the correction for Malmquist bias are shown in Fig.~\ref{fig:logdist}. As expected the distribution of log-distance ratios is generally Gaussian, with typical scatter $0.074$ and $0.085$ before and after the correction for Malmquist bias. The average error given to the log-distance ratios is $0.097$, which has contributions from observational errors and the intrinsic scatter in the TF relation. The latter component dominates the error budget, as can be seen comparing the mean error and the scatter. The separate measurements in the three photometric bands are highly consistent (and highly correlated as shown in Section~\ref{sec:3bandscomb}). The measurements with and without the correction for Malmquist bias are also highly consistent, with a small shift in the log-distance ratios towards zero when the Malmquist bias is removed. Overall, the typical error on the log-distance ratios, from both observational and intrinsic sources, is $\sim7\%$. Although the corresponding linear distances are non-Gaussian and their error can not be easily quantified, this roughly corresponds to a $\sim22\%$ linear distance error.

\subsubsection{Using the 2MTF distance measurements} \label{sec:3bandscomb}
Three separate distance measurements were obtained for each galaxy using the $K$, $H$ and $J$-band photometry. As the photometry in these bands comes from the same survey and each of the template relations uses the same HI line-width as a measure of the velocity dispersion, we expect these distance measurements to be highly correlated. To explore this, we calculate the cross-correlation coefficients from the data itself, averaging over all 2062 galaxies in the final 2MTF sample. As expected the cross-correlation coefficients between the three bands are very high, $\{\rho_{KH},\rho_{KJ},\rho_{HJ}\} = \{0.983,0.981,0.993\}$ and $\{0.986,0.978,0.985\}$ for the measurements with and without the Malmquist bias correction respectively. 

We test that the correlation coefficients for the 2MTF dataset do not vary as a function of redshift by computing them in four equally spaced redshift bins, and do not vary when we look at galaxies above and below a declination of $-40^{\circ}$. In all cases we find negligible differences ($<10\%$ of the statistical error) between the cross-correlation coefficients compared to the error on the coefficients themselves, estimated using bootstrap sampling with replacement.

As the cross-correlation coefficients are so high, treating these measurements as independent and combining them is certainly incorrect. Instead we perform separate fits of the velocity power spectrum for each of the bands, and also use the single most precise measurement for each galaxy. As the measurements are so correlated the choice of which measurement to use is essentially arbitrary, so choosing the one with the smallest error is a reasonable choice.

When using the measurements to fit the velocity power spectrum, we reduce the effect of outliers by removing galaxies that lie greater than $4\sigma$ from the mean logarithmic distance. Based on the number of galaxies within the 2MTF sample, we do not expect any to lie greater than $4\sigma$ from the mean value, however there are some small number that do, likely due to systematics in the measurement of the photometry or HI line width, galaxies with unusual properties, or due to underestimation of the statistical error. Our `clipping' approach is justified as we expect all galaxies to be drawn from the same underlying distribution and their peculiar velocities to all be driven by the same growth rate of structure.

\begin{figure}
\centering
\includegraphics[width=0.5\textwidth]{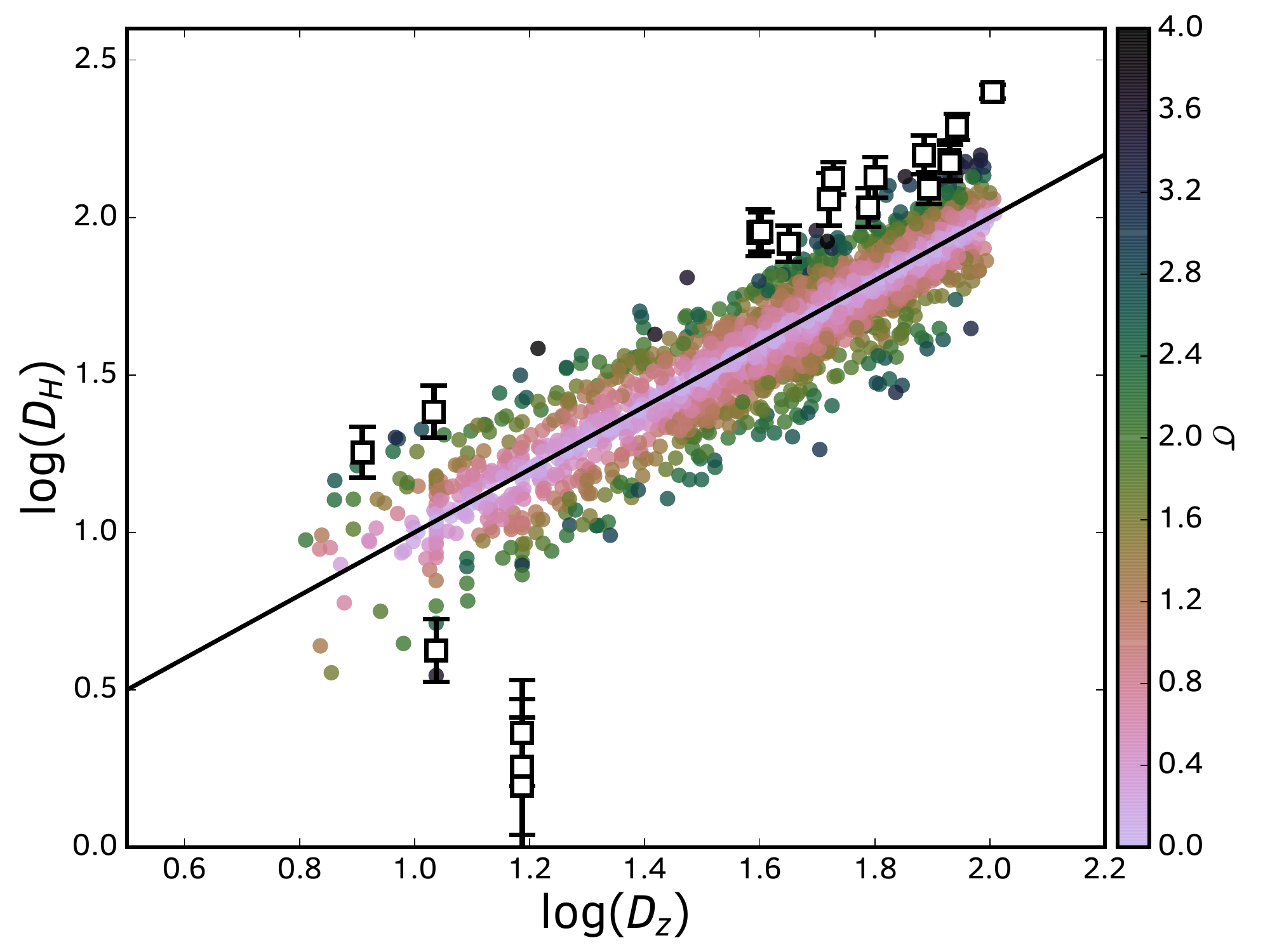} \\
  \caption{A plot of the measured distance of each galaxy from the Tully-Fisher relationship against the distance inferred from its redshift. For each galaxy we plot the Malmquist bias corrected distance with the smallest error. The color of each point signifies the deviation from the mean logarithmic distance ratio in units of the standard deviation. The open squares are points removed by the $4\sigma$ clipping.}
  \label{fig:clipping}
\end{figure}

The number of galaxies removed by the $4\sigma$-clipping is given in Table~\ref{tab:clipping}. As an example of the galaxies that we remove during this procedure, Fig.~\ref{fig:clipping} shows the logarithmic distance of each galaxy as measured from the Tully-Fisher relation against that inferred from it's redshift using our fiducial cosmology. For each galaxy we plot the measurement with the correction for Malmquist bias and the minimum error. We expect galaxies to lie about the 1:1 line, with some scatter due to their peculiar velocities. The $4\sigma$-clipping removes obvious outliers from this relation which would bias our measurement of the growth rate.

\begin{table}
\caption{The number of galaxies removed due to the $4\sigma$-clipping for the three individual bands and when using the ``minimum error'' measurement for each galaxy, and for measurements with and without the correction for Malmquist bias. The total number of galaxies before clipping is 2062.}
\centering
\begin{tabular}{l|cccc} \hline
& K-band & H-band & J-band & ``Minimum Error'' \\ \hline
With Malmquist & \multirow{2}{*}{15} & \multirow{2}{*}{17} & \multirow{2}{*}{14} & \multirow{2}{*}{19} \\
bias correction & & & & \\ \hline
Without Malmquist & \multirow{2}{*}{11} & \multirow{2}{*}{13} & \multirow{2}{*}{10} & \multirow{2}{*}{14} \\
bias correction & & & & \\ \hline
\end{tabular}
\label{tab:clipping}
\end{table}

\subsection{Mock surveys} \label{sec:mocks}
In order to test the robustness of our method for fitting the velocity power spectrum, we generate a set of mock 2MTF datasets. The known input cosmology of a simulation allows us to test the deviation of our best fit model from the input and quantify the expected systematic errors in our measurement from the data. For this purpose we start with a $z=0$ dark matter field from the SURFS simulation suite (Elahi et al., in preparation) which contains $2048^{3}$ particles in a box of $900\mpcoh$ on a side. This simulation has a mass resolution of $\sim7.4\times10^{9}h^{-1}\mathrm{M_{\odot}}$. The cosmology of the SURFS simulation matches our fiducial cosmology based on \cite{Planck2016}. This then allows us to test that our method is robust to the underlying cosmology. The simulation is run using the {\sc GADGET-2} simulation code \citep{Springel2005}. Halos and subhalos are identified from the dark matter field using the code {\sc VELOCIraptor} \citep{Elahi2011}. This code first performs a 3D Friends-of-friends algorithm \citep{DavisM1985} on the dark matter, linking together particles based on their spatial proximity, before identifying substructures using a 3D phase-space Friends-of-friends algorithm. This second step allows it to identify substructures which are dynamically distinct from the smooth, mean background of the parent halo. Halos and subhalos are identified using a minimum of 20 particles, hence the Friends-of-friends mass resolution for a halo in our simulation is $\sim1.5\times10^{11}h^{-1}\mathrm{M_{\odot}}$. For more details on the halo catalogues see Elahi et al., (in preparation). 

We expect the 2MTF sample, consisting of large, bright spiral galaxies, to reside primarily in large mass halos. For the 2MTF mocks, we simply require a simulation with a large enough volume to capture the impact of long wavelength modes on our measurements within the 2MTF survey volume and with high enough resolution that we accurately recover the typical halos in which a 2MTF galaxy resides. The volume of our simulation is much larger than the maximum extent of the 2MTF data and, given the method we use to generate our 2MTF mocks in the following sections, only $10\%$ of our mock galaxies are found in halos with less than 50 particles, which are reasonably well converged when compared to other, higher resolution simulations in the SURFS suite. Hence we consider our mocks robust to the volume and mass resolution of the simulation. 

\subsubsection{Populating the halos with galaxies}

From the catalogue of halos and subhalos, we produce a suitably realistic population of 2MTF galaxies using the Subhalo Abundance Matching technique (SHAM; \citealt{Conroy2006}). We place a single object at the centre of each halo/subhalo with a luminosity drawn from the late-type $K$-band luminosity function of \cite{Kochanek2001}. The abundance matching is performed by drawing a number of luminosities equal to the number of halos/subhalos, rank ordering this list alongside the maximum circular velocity of each structure, and assigning these one-to-one, such that the halo with the highest circular velocity contains the largest luminosity, and so on. 

The luminosity function from \cite{Kochanek2001} is a Schechter function \citep{Press1974} with $M^{*}_{K} = -22.98$, $\alpha=-0.87$ and $\bar{n}_{K}=1.01\times10^{-2}\hompcV$ fit to photometry from 2MASS, which is also the parent sample from which all 2MTF galaxies are drawn. \cite{Springob2016} found that the luminosity function of the 2MTF data has a slightly steeper slope than that of \cite{Kochanek2001} ($\alpha=-1.1$ rather than $\alpha=-0.87$), likely due to the additional morphological cuts placed on the 2MTF sample. Whilst this difference in slope is $\sim2\sigma$, the effect (when changing the slope but keeping the other parameters fixed) on the relative assignment of luminosities to halos is small and as \cite{Springob2016} do not evaluate the normalisation of their luminosity function we opt to use the fit from \cite{Kochanek2001} instead.

\subsubsection{Survey selection effects}

Once we have populated a simulation with galaxies we then place a set of observers in the box and for each observer apply a set of selection effects to reproduce the 2MTF data. We use eight observers equally spaced in the full simulation volume. Even though these mocks are drawn from the same simulation, we treat them as independent in this work. The 2MTF survey extends for roughly $\sim 100\mpcoh$ in each direction, so we can place the eight mocks such that they are always more than $250\mpcoh$ from each other in any direction. Hence we expect the different mocks to only be correlated on the largest scales, and even then only slightly.
For each observer we \textit{reproduce} the 2MTF selection function by doing the following:
\begin{enumerate}
\item{Convert the comoving distance and peculiar velocity of each galaxy to a redshift in the observer's rest frame using Eq.~\ref{eq:redshift}}
\item{Remove all objects with $cz<600\,\mathrm{km\,s^{-1}}$ and $cz>10,000\,\mathrm{km\,s^{-1}}$.}
\item{Calculate the \textit{apparent} $K$-band magnitude of each object seen by the observer and apply a cut of $K_{\mathrm{mag}}<11.25$.}
\item{Convert each remaining galaxy's cartesian coordinates to a Right Ascension and Declination. From these compute the galactic latitude, $b$, and remove all galaxies with $|b| < 5^{\circ}$. This roughly mimics the survey `mask' created by the Zone of Avoidance around the Galactic plane.}
\item{Finally, subsample the galaxies so that they fit the redshift distribution of the data. We subsample the regions of the sky above and below a Declination of $\delta=-40.0^{\circ}$ separately. For each of these regions we subsample based on a smooth spline fit to the number of objects in redshift bins (as opposed to the true number of objects in the 2MTF sample), so as not to remove too much of the naturally occurring substructure along the line of sight in each mock.}
\end{enumerate}
The number density of objects in the mock surveys for the two distinct sky areas is compared to the 2MTF data in Fig.~\ref{fig:nz}. We see a good broad agreement between the two, with small differences occurring due to the natural large scale structure along the line-of-sight in the 2MTF data. The number density of objects below $\delta=-40.0^{\circ}$ is noticeably less than at higher latitudes for all redshifts due to the different telescopes available to use to obtain the HI line-widths in these two regions.

\subsubsection{Error assignment}

We assign measurements to each of the mock galaxies in a way which also reflects the data. Log-distance ratios are calculated for each of our mock galaxies based on their measured redshifts with respect to the observer and their true comoving distance. Gaussian errors are then assigned to these mock galaxies by fitting the errors in $\Delta d$ from 2MTF as a function of redshift. We use the minimum error across the three photometric bands, as described in Section~\ref{sec:3bandscomb}. There are two sets of 2MTF $\Delta d$ measurements for each photometric band, and hence for our ``minimum error'' measurements; those without any correction for Malmquist bias, and those after a correction has been applied based on \cite{Springob2016}. We fit separate relations to the errors on the $\Delta d$ measurements for both of these cases and find best-fit relations of $\sigma(\Delta_{d}) = 0.118(\pm0.001)-1.469(\pm0.048)z$ and $\sigma(\Delta_{d*}) = 0.116(\pm0.001)-1.322(\pm0.049)z$ for the errors with and without the correction for Malmquist bias respectively. The uncertainties for these fits are calculated using 10000 iterations of bootstrap resampling with replacement. 

We also find considerable scatter around this best-fit model. To improve the realism of our mocks we quantify this scatter and incorporate it into our error assignment for the mock galaxies. We split the data into redshift bins of width $cz=1,000\,\mathrm{km\,s^{-1}}$ and calculate the standard deviation of the data about about the best-fit, which we call $\epsilon$ to avoid confusion with the error of the log-distance ratio itself. We then fit this as a function of redshift too and find $\epsilon = 0.020(\pm0.001)-0.236(\pm0.059)z$ and $\epsilon* = 0.020(\pm0.001)-0.211(\pm0.058)z$ for the scatter in the errors with and without the correction for Malmquist bias respectively. Overall, to test our assumption that the errors on the 2MTF data can be well represented by a Gaussian distribution with mean $\sigma(\Delta_{d})$ and standard deviation $\epsilon$ (which vary with redshift) we perform a two-sample Kolmogorov-Smirnov test. For each redshift bin we draw a number of samples equal to the number of data points from the corresponding Gaussian. We repeat this multiple times and find that the data supports our assumption\footnote{More specifically, performing this test 1000 times we find p-values greater than 0.1 in 935 and 647 cases for the data/fits with and without the correction for Malmquist bias respectively.}.

\begin{figure}
\centering
\includegraphics[width=0.5\textwidth]{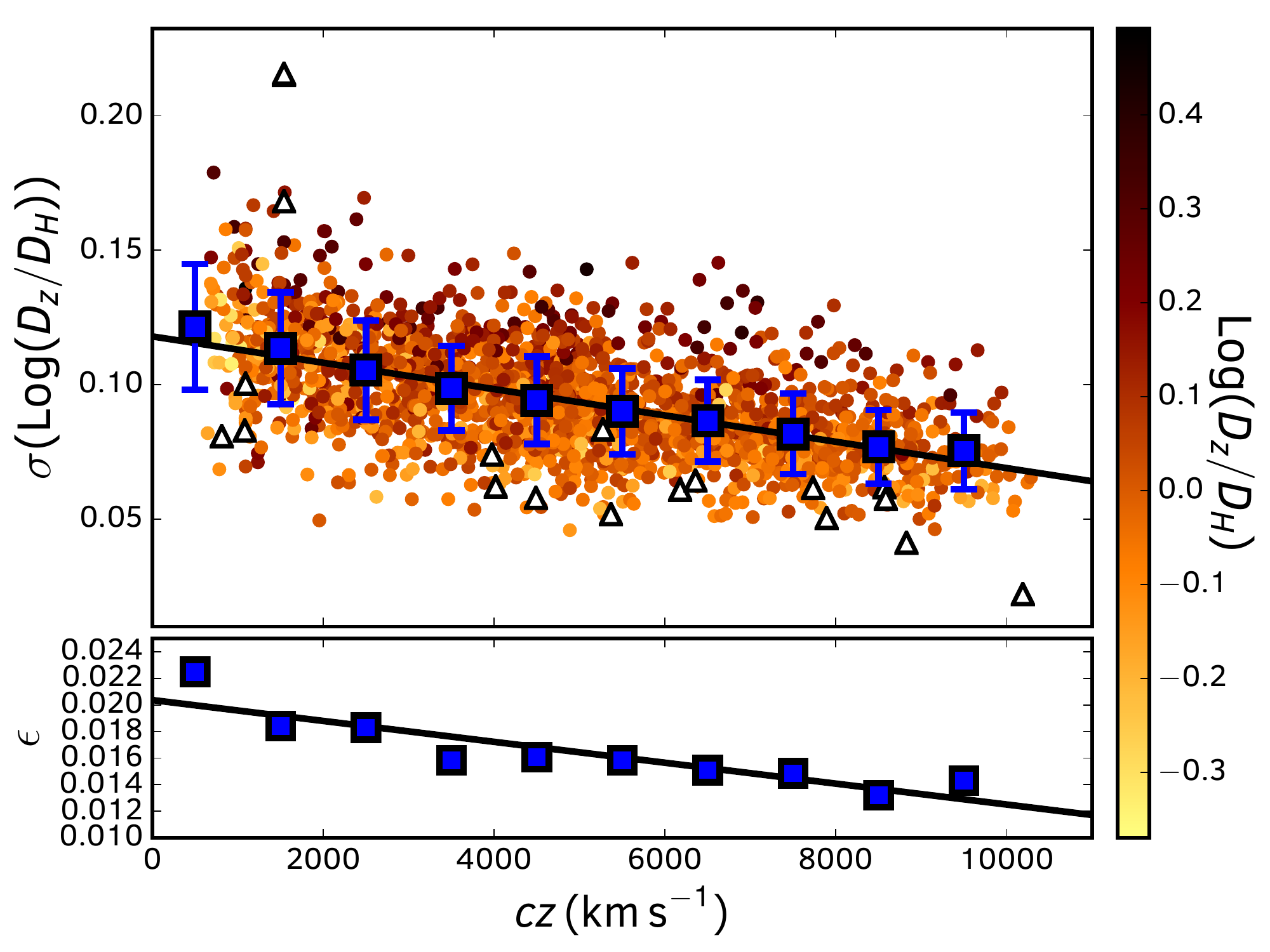} \\
  \caption{The errors in the log-distance ratios of the 2MTF data (circle points), colour-coded based on the log-distance ratio itself, and plotted alongside the best-fit model (black line) used to generate the errors in the mocks. The blue squares show the mean and standard deviation of the data in bins of width $cz=1,000\,\mathrm{km\,s^{-1}}$. The bottom panel shows these standard deviations themselves as a function of redshift alongside a best-fit model, which highlights how the scatter about our best-fit for the errors also has a slight trend with redshift. We use a combination of these two best-fit lines to estimate the mean and variance of the Gaussian PDF from which the error on a galaxy with some redshift is generated. The colour-coding of the points suggests that any correlation between the errors and the log-distance ratios is small, and so is not accounted for in this work. The open triangles show points removed by $4\sigma$ clipping which are not used in the fits.}
  \label{fig:errs}
\end{figure}

This process is demonstrated in Fig.~\ref{fig:errs}, where we plot the Malmquist-bias corrected errors in the 2MTF data as a function of redshift, alongside the best-fit relationship. We also plot the mean and variance of the data in redshift bins, and the scatter and subsequent redshift as a separate panel. The 2MTF data points are coloured based on their actual log-distance ratio to identify any additional trends as a function of this measured variable. We find some small evidence for a trend with log-distance ratio, with larger ratios corresponding to larger errors, however this is much less significant than the trend in the mean and scatter of the errors as a function of redshift and so we do not account for this. We also show the points that were removed due to the $4\sigma$ clipping described in Section \ref{sec:3bandscomb}, which are \textit{not} included when we fit the errors as a function of redshift.

From Fig.~\ref{fig:errs} we see that the errors on log-distance ratios decrease with redshift. This is due to an implicit selection function within the 2MTF; only the most HI luminous galaxies, with the largest rotation widths, can be detected at high redshift. The scatter in the TF relation was found to be well correlated with the HI line width, with higher line-widths having lower scatter \citep{Hong2014}. Hence galaxies at higher redshift have lower intrinsic scatter about the mean TF relation and smaller errors.

After obtaining the best-fit relationships the error on each mock galaxy is generated as a random number drawn from a Gaussian distribution with mean $\sigma(\Delta d)$ and standard deviation $\epsilon$. The errors are then used to perturb the measurements for each galaxy from its true value, again assuming a Gaussian distribution.

\section{Theory and modelling} \label{sec:model}

\subsection{Gaussian Theory}

In order to extract a measurement of the velocity power spectrum from the 2MTF data we use the method of \cite{Macaulay2012} and \cite{Johnson2014}. All modelling is done at $z=0$, which is close to the mean redshift of the 2MTF data. Our measurements of the velocity field are in the form of line-of-sight peculiar velocities $s(\bx) = \bv(\bx)\cdot\hat{r}$. Under the assumption that the velocities $\bv(\bx)$ are drawn from a Gaussian distribution with zero mean, the probability of observing a set of \textit{line-of-sight} peculiar velocities $\boldsymbol{s}$ is given by
\begin{equation}
\mathcal{L(\boldsymbol{\theta})} = \frac{1}{2\pi|\bC(\boldsymbol{\theta})|}\mathrm{exp}\left(-\frac{1}{2}\boldsymbol{s}^{T}\bC(\boldsymbol{\theta})^{-1}\boldsymbol{s}\right).
\label{eq:likelihood}
\end{equation} 
The velocity covariance matrix $\bC$ for this set of observations depends on the underlying cosmological model and parameters $\boldsymbol{\theta}$, and the relative positions of the galaxies in the data vector $\boldsymbol{s}$. For two galaxies, $i$ and $j$, we have $C_{ij} = \langle s_{i}(\bx_{i}) s_{j}(\bx_{j}) \rangle$. From Eq.~\ref{eq:likelihood} we can calculate the likelihood of measuring our set of peculiar velocities given some undelying cosmological model. Using Bayes' theorem, we can then calculate the posterior distribution of a set of cosmological parameters given our peculiar velocity dataset, the likelihood in Eq.~\ref{eq:likelihood}, and the priors and method given in Section~\ref{sec:freeparameters}.

Theoretical modelling of the correlations between velocities in disparate locations is typically done in terms of the velocity power spectrum $P_{vv}(k,a)$, or the velocity-divergence power spectrum $P_{\theta\theta}(k,a)$. The relationship between the two on linear scales at $z=0$ is (e.g., \citealt{ColesLucchin}; Eq.18.1.13)
\begin{equation}
P_{vv}(k) = \left(\frac{H_{0}f(k)}{k}\right)^{2}P_{\theta\theta}(k).
\label{eq:pkvel}
\end{equation}
This relationship between the velocity power spectrum and velocity divergence power spectrum follows from Eq.~\ref{eq:continuity}. Our likelihood evaluation requires the covariance matrix in real-space, but we can write this in terms of the velocity power spectrum by first using Fourier transforms to relate it to the peculiar velocities in $k$-space,
\begin{equation}
C_{ij}(\bx_{i},\bx_{j}) = \int \frac{d^{3}k}{(2\pi)^{3}}e^{i\bk\cdot\bx_{i}} \int \frac{d^{3}k'}{(2\pi)^{3}}e^{-i\bk'\cdot\bx_{j}} \langle s_{i}(\bk)s^{*}_{j}(\bk') \rangle,
\end{equation}
then substituting the line-of-sight PVs for the underlying velocities and writing their variance in terms of the velocity power spectrum. Separating the resulting integral into radial and angular components we find,
\begin{align}
C_{ij}(\bx_{i},\bx_{j}) = \frac{H_{0}^{2}}{2\pi^{2}}\int dk f^{2}(k)P_{\theta\theta}(k,a)W(\bx_{i},\bx_{j},k),
\label{eq:cov2}
\end{align}
where
\begin{equation}
W(k,\bx_{i},\bx_{j}) = \int \frac{d^{2}k}{4\pi}e^{i\bk\cdot(\bx_{i}-\bx_{j})}(\hat{x}_{i}\cdot\hat{k})(\hat{x}_{j}\cdot\hat{k}).
\label{eq:win1}
\end{equation}
\cite{Ma2011} give an analytic expression for the window function in terms of the comoving distance to the two galaxies, their radial separation $A_{ij}=|\br_{i}-\br_{j}|$, and the angle between them $\alpha_{ij} = \mathrm{cos}^{-1}(\hat{x}_{i}\cdot\hat{x_{j}})$,
\begin{align}
W(k,\bx_{i},\bx_{j}) &= 1/3[j_{0}(kA_{ij}) - 2j_{2}(kA_{ij})]\mathrm{cos}(\alpha_{ij}) \notag \\
&+ A_{ij}^{-2}j_{2}(kA_{ij})x_{i}x_{j}\mathrm{sin}^{2}(\alpha_{ij}).
\label{eq:win2}
\end{align}

Given a sample of galaxies with measured positions, redshifts and peculiar velocities, we can:
\begin{enumerate}
\item{Adopt a given cosmological model to convert the galaxy coordinates to cartesian coordinates, and evaluate the velocity divergence power spectrum and the necessary prefactors in Eq.~\ref{eq:cov2}}
\item{Compute the covariance matrix for all possible galaxy pairs. Evaluating the integral in Eq.~\ref{eq:cov2} requires choosing appropriate integration limits. Theoretical models of the velocity divergence power spectrum will break down at some non-linear scale. Including these scales in the integral can bias results, so the range of scales we choose to  integrate over and fit against must be chosen appropriately.}
\item{Calculate the likelihood for the cosmological model based on the covariance matrix and the peculiar velocity measurements.}
\end{enumerate}
Iterating over these steps allows us to evaluate our posterior.

In practice, there are a few caveats with this approach. We first require a way to incorporate measurement errors into our likelihood calculation, which in most applications is not trivial. We also need a method to calculate the velocity divergence power spectrum that is accurate to the scales we wish to fit against. If this is not available, we can suppress non-linearities in the data and use a more linear model. Ideally, we try to achieve some balance between these two options. Finally, we need to include marginalisation over the effects of zero-point offsets, or a monopole, in the peculiar velocity measurements. Methods to include these are summarised in the following sections.

\subsection{Measurement Errors}

Measurements of the peculiar velocities of galaxies are subject to considerable statistical errors, which must be incorporated into our likelihood analysis. As long as the distribution of the measured values about the true underlying peculiar velocities can be written as some probability distribution function (PDF), we can calculate the likelihood of measuring a particular configuration of peculiar velocities as a convolution between this PDF and the multivariate Gaussian likelihood arising from the theory, Eq.~\ref{eq:likelihood}. This statement is true in general, however writing an analytic formula for the result of this convolution is particularly difficult (or may even yield an answer with no closed form) unless the measurement errors are also Gaussian. In this case the likelihood of interest is a convolution between the Gaussian theory and a multivariate Gaussian describing the measurement errors, the result of which is simply another Gaussian, albeit with a different mean and variance. 

For Gaussian distributed errors, if we assume that the measurements are unbiased (in that the measurement PDF is centred on the true underlying value) and independent (such that the covariance matrix of this PDF is diagonal) we find that the joint likelihood including both the theory and measurement errors is simply a modified version of Eq.~\ref{eq:likelihood},
\begin{equation}
\mathcal{L(\boldsymbol{\theta})} = \frac{1}{2\pi|\boldsymbol{\mathsf{\Sigma}}(\boldsymbol{\theta})|}\mathrm{exp}\left(-\frac{1}{2}\boldsymbol{s}^{T}\boldsymbol{\mathsf{\Sigma}}(\boldsymbol{\theta})^{-1}\boldsymbol{s}\right),
\label{eq:likelihood2}
\end{equation}
where
\begin{equation}
\Sigma_{ij} = C_{ij} + \sigma^{2}_{i}\delta_{ij},
\end{equation} 
and $\sigma_{i}$ is the error on the measurement for galaxy $i$. 

Unfortunately, the errors on peculiar velocities are generally log-normal, not Gaussian, and possess significant bias and skewness. One way of dealing with this could be to write down a form of the Non-Gaussian measurement PDF that can still be convolved analytically. Though of great interest, this is a significant undertaking and beyond the scope of this work\footnote{To the best of the author's knowledge a closed solution for the convolution between a multivariate Gaussian and a second multivariate PDF does not exist for Lognormal distributions. However, certain classes of multivariate Skew-Normal distribution (see e.g. \cite{Azzalini2009}) can be convolved with a Gaussian PDF and result in another Skew-Normal distribution. It is possible that this could then be used to evaluate the likelihood for a given dataset.}. A simpler method is to perform a change of basis to a variable that has a Gaussian distribution. \cite{Watkins2015} and \cite{Johnson2014} present two such transformations, which will be covered separately.

In either case, in this work the error on each galaxy is assumed to come from a combination of observational uncertainty $\sigma_{obs,i}$, which also includes the scatter in the Tully-Fisher relation, and a stochastic noise arising from non-linear motions $\sigma_{*,i}$. This stochastic noise is typically on the order of $200-300\mathrm{km\,s^{-1}}$ \citep{Masters2006, Scrimgeour2016}, but we include it as a free parameter in our likelihood evaluation rather than fixing it to a specific value. Hence the total error on each galaxy is $\sigma_{i}^{2} = \sigma_{obs,i}^{2} + \sigma_{*,i}^{2}$.

\subsubsection{Gaussian estimator for peculiar velocities}

\cite{Watkins2015} (WF15) present an estimator for the peculiar velocity of a galaxy based on the logarithmic distance ratio that is unbiased for galaxies with peculiar velocities much less than their redshifts. They show that their estimator 
\begin{equation}
s_{i} = \frac{cz_{m,i}}{1+z_{m,i}}\mathrm{ln}(10)\Delta d_{i},
\label{eq:wf15}
\end{equation}
also has a Gaussian PDF. $z_{m,i}$ is a corrected redshift for galaxy $i$ which includes the effects of cosmic acceleration. \cite{DavisT2014} give an expression for this in terms of the deceleration parameter, $q_{0}$
\begin{equation}
z_{m} = z[1 + 1/2(1-q_{0})z - 1/6(2-q_{0}-3q_{0}^{2})z^{2}].
\end{equation}
For observational errors in the log-distance ratio a similar transformation can be applied such that $\sigma_{i}^{2} = (cz_{m}/(1+z_{m}))^{2}(\mathrm{ln}\,10)^{2}\sigma_{obs,i}^{2} + \sigma_{*,i}^{2}$.

All the galaxies in the 2MTF sample have redshifts larger than $600\mathrm{km\,s^{-1}}$ which is expected to be equal or greater than the typical peculiar velocity for a galaxy. However, while this estimator should be applicable to the 2MTF sample, it may be inaccurate for the lowest redshift galaxies, or a small number of galaxies with extraordinarily large PVs. This will be tested using the mock catalogues in Section~\ref{sec:tests}.

\subsubsection{Theory for magnitude fluctuations}

An alternative variable we can use is the fluctuation in apparent magnitude caused by the peculiar motion of an object, $\delta m$, which is that used by \cite{Johnson2014}. The relation between the log-distance ratio and a magnitude fluctuation is the same as that given for absolute magnitudes in Eq.~\ref{eq:mag}. Using this variable requires us to rewrite the equation for the covariance matrix for the velocities in terms of apparent magnitude fluctuations.

The derivation of how an object's peculiar velocity changes its observed magnitude has been shown many times in the literature (see, for example \citealt{Hui2006,DavisT2011,Johnson2014,Huterer2015}) and so will not be repeated here. Ultimately, ignoring effects beyond first order in perturbation theory, and ensuring that we work in the CMB frame, the change in apparent magnitude induced by an object's peculiar velocity is
\begin{equation}
\delta m = -\frac{5}{\mathrm{ln}\,10}\frac{1+z}{H(z)\chi(z)}s.
\label{eq:magfluc}
\end{equation}
$H(z)$ is again the Hubble parameter, this time at the redshift of the galaxy, and $\chi(z)$ is the comoving distance to this galaxy.

Combining Eqs.~\ref{eq:mag} and \ref{eq:magfluc}, the covariance matrix for log-distance ratios between two galaxies is then
\begin{equation}
C^{\Delta d}_{ij} =  \left(\frac{1}{\mathrm{ln}\,10}\right)^{2}\left(\frac{1+z_{i}}{H(z_{i})\chi(z_{i})}\right)\left(\frac{1+z_{j}}{H(z_{j})\chi(z_{j})}\right)C_{ij},
\label{eq:magfluccov}
\end{equation}
where $C_{ij}$ is the covariance matrix as defined in Eq.~\ref{eq:cov2}. When using this parameterisation we use $\boldsymbol{\delta m}$ as our data vector in Eq.~\ref{eq:likelihood2} rather than $\boldsymbol{s}$, and we also have to transform the variance in the measured velocities due to the stochastic noise, such that $\sigma_{i,\Delta d}^{2} = \sigma_{obs,i}^{2} + (\mathrm{ln}\,10)^{-2}((1+z_{i})/(H(z_{i})\chi(z_{i}))^{2}\sigma_{*,i}^{2}$.

The derivation of the prefactor to convert the covariance matrix for peculiar velocities into that of log-distance ratios relies on a Taylor expansion in $\mathrm{log}(1+x)$, where $x$ depends on the peculiar velocity and distance of each galaxy. This expansion only converges for $x<<1$, and investigation of this term shows that this may not be satisfied for nearby galaxies with large peculiar velocities. As with the WF15 estimator, we will test the validity of this method using our mock catalogues in Section~\ref{sec:tests}.

\subsection{Modelling the velocity divergence power spectrum}

In this work, we use a model velocity divergence power spectrum generated using the implementation of two-loop Renormalised Perturbation Theory (RPT; \citealt{Crocce2006a, Crocce2006b, Crocce2008}) found in the {\sc copter} numerical package \citep{Carlson2009}. This takes as input a fiducial cosmology and a corresponding linear matter transfer function normalised to unity at $k\rightarrow 0$, which we generate using {\sc camb} \citep{Lewis2000, Howlett2012}. \cite{Carlson2009} found that two-loop RPT was able to recover the real-space density-density, density-velocity and velocity-velocity power spectra of a redshift $z=0$ $\Lambda$CDM universe to within $1\%$ up to $k=0.08\hompc$ and $\sim 8, 10$ and $15\%$ respectively up to $k=0.2\hompc$. Improved accuracy could be obtained using emulators of the power spectrum (e.g., \citealt{Heitmann2014,Agarwal2014}), however these currently only exist for the matter power spectrum, not the velocity divergence power spectrum. As the accuracy of our theoretical power spectrum breaks down as we include more non-linear scales in the model, so we must either restrict our fits to scales where these inaccuracies remain small, possibly sacrificing constraining power, or suppress non-linearities in the data.

On top of this, our fitting of the velocity power spectrum requires us to estimate the distance to each galaxy in order to compute the window function and in turn the covariance matrix. Although we have measurements in the 2MTF data of the distance to each galaxy, these contain considerable statistical error. Instead we convert the measured redshift of each object to a comoving distance, which contains contributions from the peculiar velocities themselves. This gives rise to non-linear Redshift Space Distortions (RSD) in our model, which would not occur if we used the true distance to each galaxy. As with the numerical inaccuracies in the modelling, these can be suppressed or they can be included in the model itself. 
  
In this section we will present two methods to overcome these non-linearities: Firstly, by including an additional free parameter to model non-linear RSD on small scales, and secondly by gridding the data to smooth out non-linear effects. This latter method was adopted by \cite{Johnson2014}. As with our choice of variable in the previous section, we will justify the choice of method used to model the velocity power spectrum when fitting the 2MTF data by testing these different schemes using the mock catalogues.

\subsubsection{Including non-linear RSD}

Our first method to include non-linear effects in the model is to include an additional damping of the velocity power spectrum of small scales, mimicking that caused by non-linear RSD. We adopt the same parameterisation used by \cite{Koda2014} and \cite{Howlett2017} in their model of the velocity power spectrum,
\begin{equation}
P_{vv}(k) = \left(\frac{H_{0}f(k)}{k}\right)^{2}D^{2}_{u}(k,\sigma_{u})P_{\theta\theta}(k).
\end{equation}
Comparing this with Eq.~\ref{eq:pkvel} highlights the inclusion of the non-linear damping term $D_{u}(k,\sigma_{u})$, which contains the additional free parameter $\sigma_{u}$. The functional form
\begin{equation}
D_{u}(k,\sigma_{u}) = \mathrm{sinc}(k\sigma_{u})
\end{equation}
was found by \cite{Koda2014} to be a good match to halos in a variety of mass bins from the GiggleZ simulation \citep{Poole2015}. In all cases this model of the velocity power spectrum provided a good fit for $k\le0.2\hompc$, with the best-fit value of $\sigma_{u}$ showing some slight dependence on halo mass between the values of $\sigma_{u}=13.0-15.5\mpcoh$, with higher values corresponding to more massive halos, and hence more non-linear damping of the velocity power spectrum.   

Although small scale damping due to RSD is typically a function of both the scale $k$ and the angle between the $\bk$-vector and the observer's line of sight (typically denoted $\mu$), no evidence for such a dependency was found by \cite{Koda2014}. Because this damping model only depends on scale, including it in the covariance matrix is trivial. In this case, Eq.~\ref{eq:cov2} becomes
\begin{multline}
C^{RSD}_{ij}(\bx_{i},\bx_{j}) = \frac{H^{2}_{0}}{2\pi^{2}}\int dk f^{2}(k)P_{\theta\theta}(k,a) \\
\quad D^{2}_{u}(k,\sigma_{u})W(\bx_{i},\bx_{j},k).
\end{multline}

\subsubsection{Gridding the data}

Alternatively, or in addition to including the non-linear damping caused by RSD in the model, we can suppress non-linearities in the data by smoothing the measurements on some grid and calculating the theoretical covariance at the cell centres. This has the benefit of also potentially reducing the size of the covariance matrix we need to compute from the data. This method was introduced and tested by \cite{Abate2008} and \cite{Johnson2014} and found to produce unbiased results for suitable choices of grid size and fitting range.

When gridding the data, galaxies are assigned to the grid point. The mean of their log-distance ratios is taken as the value in the cell. As we are treating these measurements as uncorrelated, we take the error in cell $i$, $\sigma_{grid,i}$, as the standard error on the mean
\begin{equation}
\sigma^{2}_{grid,i} = \frac{1}{n_{i}}\sum_{j}\sigma^{2}_{j}\Theta_{ij},
\end{equation}
where $n_{i}$ is the number of galaxies in that cell, and $\Theta_{ij} = 1$ if galaxy $j$ is in cell $i$ and $0$ otherwise\footnote{Note that this is slightly different to the definition used in \cite{Abate2008} and \cite{Johnson2014}. They use the average error in each cell, divided by $n^{1/2}_{i}$ to account for the error reduction due to averaging, i.e., $\sigma_{grid,i} = \frac{1}{n^{3/2}_{i}}\sum_{j}\sigma_{j}\Theta_{ij}$. This is only coincident with the standard error on the mean when there is a single galaxy in the cell, or the error on each galaxy is the same. This is not true for the 2MTF sample, but in practice the differences in the error in each cell when using the standard error on the mean, or when averaging the error as in \cite{Abate2008} and \cite{Johnson2014}, are small.}. The error added to each cell due to random non-linear motions is similarly calculated.

When gridding the data we are effectively suppressing non-linear power. This has to be reflected in the calculation of the covariance matrix. By taking the Fourier transform of the gridding kernel $\Theta_{ij}$, and angle-averaging this we obtain a function $\Gamma(k)$ that can be used to suppress the non-linear power spectrum when calculating Eq.~\ref{eq:cov2}. The equation for the covariance matrix of the gridded data is then
\begin{multline}
C^{\mathrm{grid}}_{ij}(\bx_{i},\bx_{j}) = \frac{H^{2}_{0}}{2\pi^{2}}\int dk f^{2}(k)P_{\theta\theta}(k,a)\Gamma^{2}(k)W(\bx_{i},\bx_{j},k),
\end{multline}
where 
\begin{align}
\Gamma(k) &= \frac{1}{4\pi}\int^{\pi}_{0}\int^{2\pi}_{0}d\theta d\phi\,\mathrm{sinc}(k_{x}) \mathrm{sinc}(k_{y}) \mathrm{sinc}(k_{z}), \notag \\
k_{x} &= (kL/2)\mathrm{sin}(\theta)\mathrm{cos}(\phi), \notag \\
k_{y} &= (kL/2)\mathrm{sin}(\theta)\mathrm{sin}(\phi), \notag \\
k_{z} &= (kL/2)\mathrm{cos}(\theta), 
\end{align}
and $L$ is the edge-length of the gridcells we are using.

The gridding of the data makes the implicit assumption that the PVs inside each cell are well-described by a continuous field. However this becomes less valid as the number of galaxies in each cell becomes small, which can in turn bias results. \cite{Abate2008} proposed and tested a correction for this, where the diagonal elements of the covariance matrix are updated as
\begin{equation}
C^{\mathrm{grid}}_{ii} \rightarrow C^{\mathrm{grid}}_{ii} + (C_{ii} - C^{\mathrm{grid}}_{ii})/n_{i},
\end{equation}
and $C_{ii}$ is the standard ungridded covariance matrix. In the limit of one galaxy in a cell, the gridded covariance matrix returns to the standard covariance matrix. No such correction was found to be necessary for the off-diagonal elements of the covariance matrix as these are negligible on small scales.

\subsection{Velocity Monopoles}
The final addition to the theoretical modelling is an analytic correction for the effect of a monopole in the velocity field caused by an offset in the zero-point of the TF relation, or local small-scale inhomogeneities in our local universe. \cite{Howlett2017} show that this acts as a shot-noise contribution to the power spectrum, but is unlikely to cause significant bias in measurements of the growth rate as this is much smaller than the typical statistical error in the distances, which also acts as shot noise. As the 2MTF survey is homogeneous and approximately full-sky, we also expect it to be less affected by a zero-point offset than other surveys. Nonetheless, \cite{Johnson2014} showed that analytically marginalising over this is trivial, and so we include this in our analysis.

If we define the velocity monopole as a constant additive term to the peculiar velocities or magnitude fluctuations of the 2MTF galaxies, drawn from a Gaussian prior with zero mean and standard deviation $\sigma_{y}$, we can analytically marginalise over the unknown monopole by modifying Eq.~\ref{eq:likelihood2} \citep{Bridle2002}. The likelihood function we calculate becomes
\begin{equation}
\mathcal{L(\boldsymbol{\theta})} = \frac{(1+\boldsymbol{I}^{T}\boldsymbol{\mathsf{\Sigma}}^{-1}\boldsymbol{I}\sigma_{y}^{2})}{2\pi|\boldsymbol{\mathsf{\Sigma}}(\boldsymbol{\theta})|}\mathrm{exp}\left(-\frac{1}{2}\boldsymbol{s}^{T}\boldsymbol{\mathsf{\Sigma}_{m}}(\boldsymbol{\theta})^{-1}\boldsymbol{s}\right),
\end{equation}
where 
\begin{equation}
\boldsymbol{\mathsf{\Sigma}_{m}}^{-1} = \boldsymbol{\mathsf{\Sigma}}^{-1} - \frac{\sigma_{y}^{2}\boldsymbol{\mathsf{\Sigma}^{-1}}\boldsymbol{I}\boldsymbol{I}^{T}\boldsymbol{\mathsf{\Sigma}^{-1}}}{1+\boldsymbol{I}^{T}\boldsymbol{\mathsf{\Sigma}^{-1}}\boldsymbol{I}\sigma_{y}^{2}},
\end{equation}
and $\boldsymbol{I}$ is a vector of ones. The expression is similar when using magnitude fluctuations as our variable. In all cases we use a value of $\sigma_{y} = 0.2$ for our prior, but find that the exact value has little effect on our results (c.f. Section~\ref{sec:tests}).

\subsection{Free parameters and application}\label{sec:freeparameters}
In this section we have detailed the theory behind the velocity power spectrum and how a sample of PVs can be used to measure this and constrain the growth rate. One caveat is that in modelling the covariance matrix the value of the growth rate is exactly degenerate with the intrinsic amplitude of the velocity divergence power spectrum, i.e., increasing the amplitude of the velocity divergence power spectrum has the same effect on the covariance matrix as increasing $f$. We can parameterise the amplitude of the power spectrum using $\sigma_{8}$, the linear matter variance in spheres of radius $8\mpcoh$. Our formalism is then sensitive to the well-known parameter combination $f\sigma_{8}$. This parameter combination is widely used in RSD studies; \cite{Song2009} show it can be used to constrain the properties of dark energy and gravity even without explicit knowledge of $\sigma_{8}$. 

On top of $f\sigma_{8}$, we also have the free parameter $\sigma_{v}$ which parameterises the non-linear velocity dispersion of the galaxies, and $\sigma_{u}$ when we include non-linear RSD in our model. We will also perform scale dependent fits of the growth rate by calculating the integral for covariance matrix using non-overlapping $k$-bins, with a different value of $f\sigma_{8}$ for each bin. This makes the assumption that the value of $f\sigma_{8}$ is constant or does not vary widely across the bin, but allows us to look for general scale-dependence in the growth rate on large and small scales.

All of our fits are obtained with Markov chain Monte Carlo (MCMC) sampling using the publicly available {\sc emcee} routine \citep{ForemanMackey2013}. We do not allow $f\sigma_{8}$ to vary in such way that we choose unphysical negative values and use flat priors of $0 \mathrm{km\,s^{-1}} < \sigma_{v} \le 1000 \mathrm{km\,s^{-1}}$ and $0.2 \mpcoh \le \sigma_{u} \le 25 \mpcoh$. This is chosen based on the fits to the GiggleZ simulation  \citep{Poole2015} by \cite{Koda2014}. Without gridding, the covariance matrix is computationally demanding, so when fitting the velocity power spectrum with the extended non-linear RSD model we pre-compute 125 covariance matrices between our prior with width $\Delta\sigma_{u} = 0.2\mpcoh$, and linearly interpolate between these for our likelihood calculation.

When quoting our results we use the maximum likelihood value and our $1\sigma$ errors are calculated from the equal likelihood bounds containing $68\%$ of the likelihood.

\section{Tests on simulations} \label{sec:tests}
To validate our fitting method before applying it to the data, we first fit our eight mock 2MTF surveys. The questions we wish to answer are:
\begin{itemize}
\item{Which Gaussian-distributed variable is better for measuring the growth rate, magnitude fluctuations, $\delta m$ (Eqs.~\ref{eq:magfluc} and \ref{eq:magfluccov}), or peculiar velocities obtained with the WF15 estimator (Eq.~\ref{eq:wf15}).}
\item{What is the maximum $k$-value, $k_{max}$, we can use when including a free parameter for non-linear RSD? When gridding the data, what combination of $k_{max}$ and grid size returns unbiased constraints?}
\item{Does the marginalisation over the zero-point change the constraints?}
\end{itemize}
These questions will be addressed in this section. We begin our tests using the $\delta m$ variable as this was shown to return unbiased fits to 6dFGSv data by \cite{Johnson2014}, and identify the method we will use to include non-linear information in our fits to the data. We then try fitting the velocity power spectrum using the WF15 estimator before finally checking for systematic effects associated with not marginalising over the zero-point. As explained in Section.~\ref{sec:mocks}, we assign errors to our simulations based on the data with \textit{and} without the correction for Malmquist bias. We find that the differences between the mock results with these two different error assignments are negligible, and so only present the case for Malmquist bias corrected measurement errors in this section.

\subsection{Fitting to non-linear scales} \label{sec:testsnonlinear}

\begin{figure}
\centering
\includegraphics[width=0.5\textwidth]{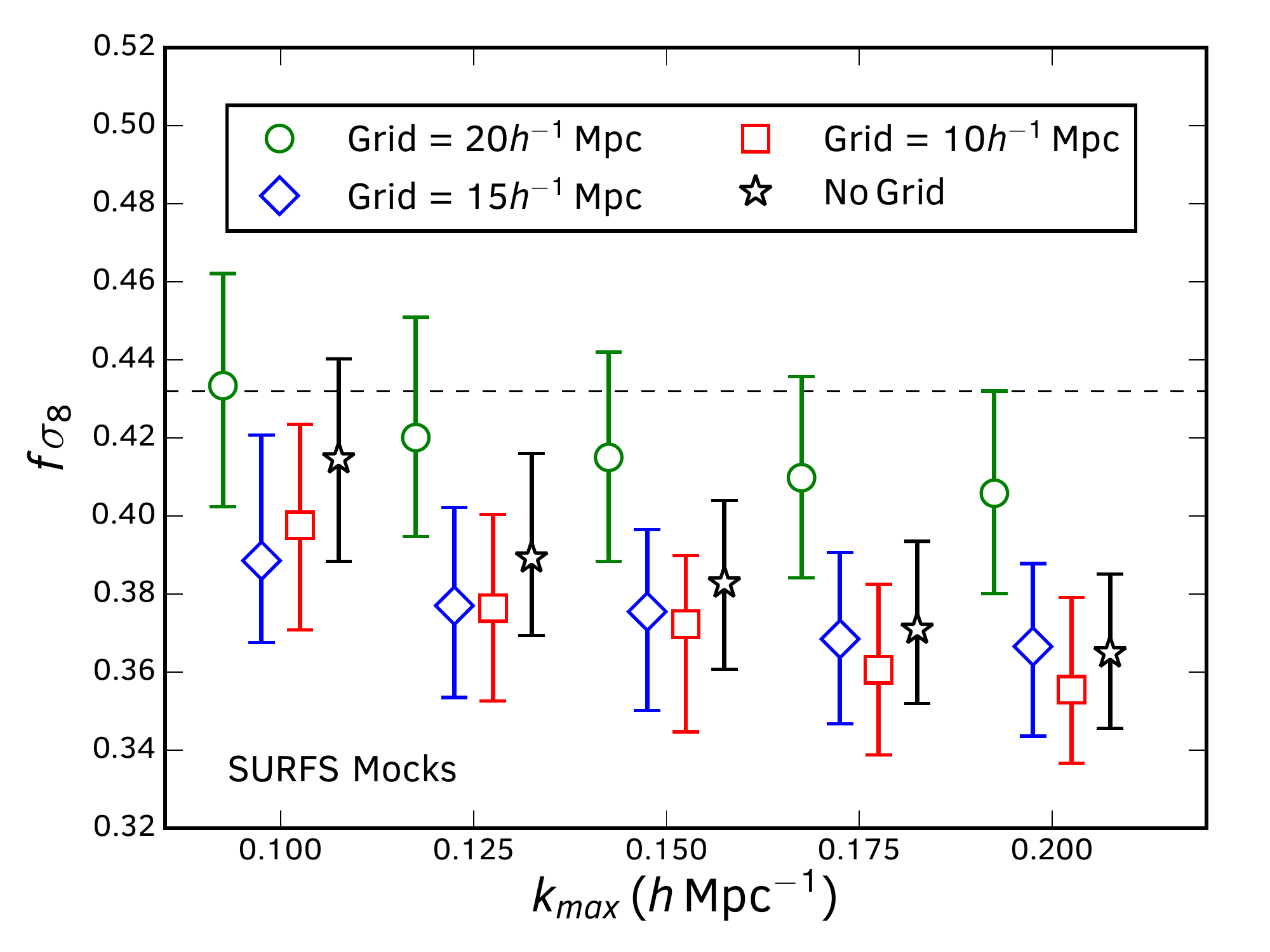} \\
  \caption{Growth rate constraints from fits to the SURFS mocks for a variety of grid sizes and $k_{max}$ values. Different symbols/colors indicate different gridsizes. The horizontal dashed line shows the correct value of $f\sigma_{8}$ based on the input cosmology of the SURFS simulations. Values of $k_{max}=[0.1, 0.125, 0.15, 0.175, 0.2]$ were used in the fits, the horizontal displacement of the points from these values is for clarity only.}
  \label{fig:gridkmax}
\end{figure}

We simultaneously fit the eight mocks for a variety of gridsizes and values of $k_{max}$, using magnitude fluctuations as our variable, assuming the true input cosmology, and marginalising over the zero point. In each case we treat the mocks as independent and evaluate the joint likelihood across all eight mocks between $k_{min}=0.007\hompc$ and $k_{max}$. As we are treating the mocks as independent, the marginalised constraints are the same as fitting each mock individually and averaging. However, this is not true for the un-marginalised constraints. Instead, combining the fits from individual mocks would require re-weighting the chain from a given mock based on the individual likelihoods from the seven other mocks. This would then have to be performed for each mock in turn, and hence would be computationally demanding. Furthermore, the results of this procedure are only robust if the parameter space is sufficiently sampled by the eight individual likelihoods.

The results are collated in Fig.~\ref{fig:gridkmax}. We find that the fits without gridding or with grid scales of $10$ and $15\mpcoh$ are consistently biased, and become more so as we go to higher $k_{max}$ and include more non-linear information in our fits. The bias is slightly worse for smaller grid sizes, but the results with both the grid sizes and without gridding are similar as in these cases a majority of cells contain a single galaxy and the data is smoothed by only a small amount. We find that smoothing on a grid size of $20\mpcoh$ is required to return unbiased constraints, and find a similar trend in the bias as a function of $k_{max}$. Overall, with this gridsize, we choose a value of $k_{max}=0.15\hompc$ to fit the data as this has a good balance between constraining power and expected systematic error. In reality, this is not much of a compromise as using a higher $k_{max}$ gives an almost negligible improvement in the statistical error compared to the obvious increase in systematic error for such large grid sizes. From the fits to the mocks we would expect a systematic error of $\sim 0.017$, much less than the expected statistical error of $\sim 0.075$ (calculated based on the statistical error from the mock fits multiplied by $\sqrt{8}$). In other words, we expect a systematic error on the data of no larger than $0.25\sigma$. The minimum grid size we require for our fits is larger than that found by \cite{Johnson2014}, but we would expect the 2MTF data, consisting of late-type galaxies and with a higher number density of nearby galaxies, to be more susceptible to non-linearities than the early-type galaxy sample of 6dFGSv.

\begin{figure}
\centering
\includegraphics[width=0.5\textwidth]{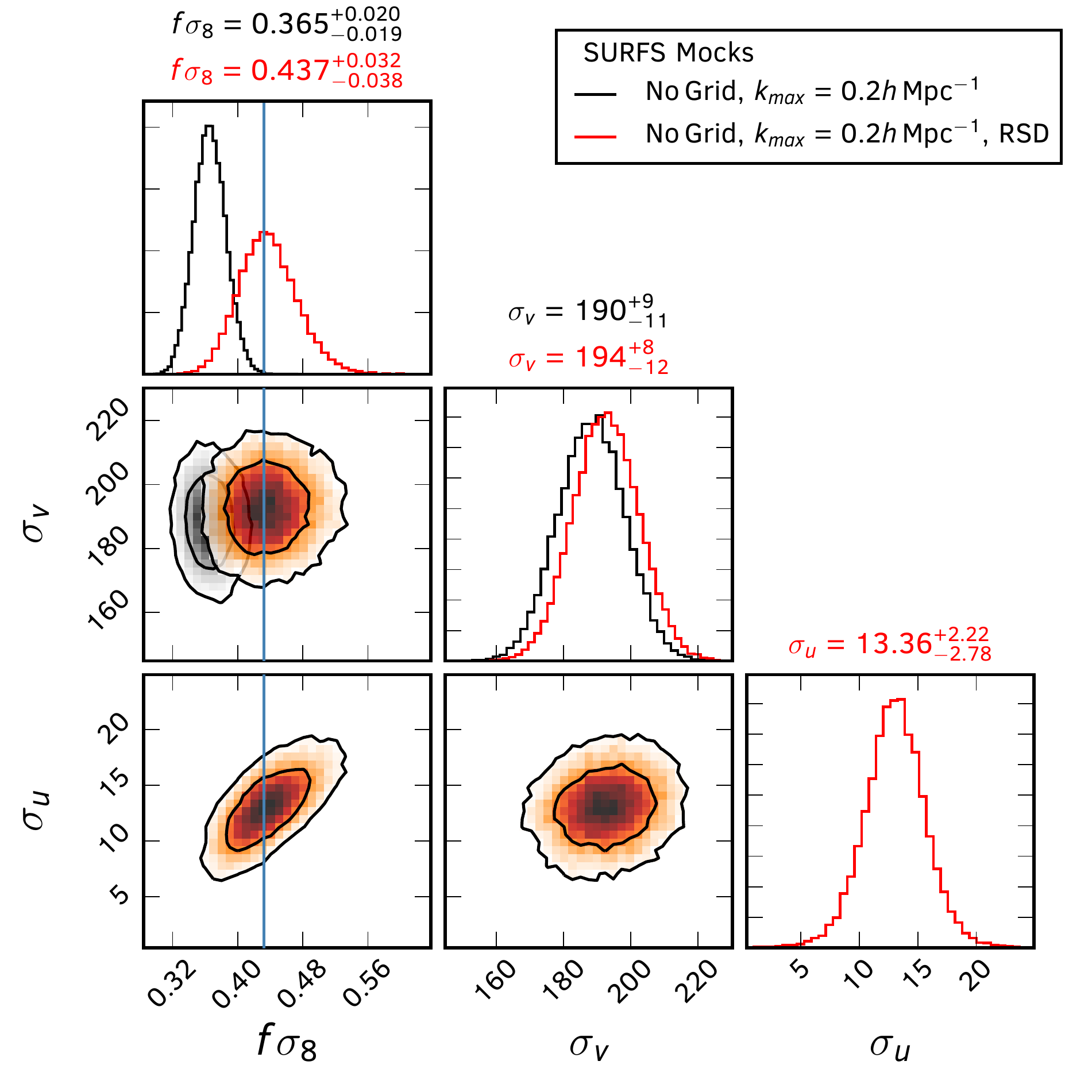} \\
  \caption{Likelihood contours and 1D marginalised histograms for fits to the SURFS mocks without gridding and using $k_{max}=0.2\hompc$. Grey contours/histograms show the results when we do not include any extension to the modelling to account for non-linearities. The resulting constraints on $f\sigma_{8}$ are biased by $0.065$, compared to the correct value for the SURFS cosmology (blue vertical line). This is larger than the expected statistical error in the data using this method ($0.055$, calculated by multiplying the statistical errors in the mocks by $\sqrt{8}$). Including non-linear RSD and the $\sigma_{u}$ parameter in the modelling (red countours/histograms) corrects for this and returns unbiased constraints, at the cost of reduced constraining power.}
  \label{fig:nogridRSDGiggleZ}
\end{figure}

We next test our second method to account for non-linearities, by including a free parameter to model non-linear RSD rather than gridding the data. When including this extension to the modelling, we find that we can obtain unbiased constraints from the SURFS mocks even when including scales down to $k_{max}=0.2\mpcoh$. This is shown in Fig.~\ref{fig:nogridRSDGiggleZ} where we plot likelihood contours and 1D marginalised histograms for the joint fits to the mocks without gridding for $k_{max}=0.2\hompc$ and with and without including the non-linear RSD modelling. As already shown in Fig.~\ref{fig:gridkmax}, the constraints without gridding the data are biased by $0.065$, which is larger than the expected statistical error in the data using this method ($0.055$, calculated as previously). Including the model extension instead returns unbiased constraints. The cost of this is reduced constraining power due to degeneracy between $f\sigma_{8}$ and the non-linear RSD parameter $\sigma_{u}$, which increases the error on the former from $5\%$ to $8\%$ in our fits to the mocks. In both cases we find consistent values for $\sigma_{v}$, both of which are within the commonly assumed range of $150-300\mathrm{km\,s^{-1}}$. The likelihood for $\sigma_{u}$ is well situated within our prior range and matches the typical values for this parameter used in \cite{Koda2014} and \cite{Howlett2017}.

\subsection{Using peculiar velocities rather than magnitude fluctuations}
We have identified the regime in which we are able to recover unbiased fits to the mocks using the $\delta m$ variable. We next look at whether the WF15 estimator of the peculiar velocity can be used instead to return similar results. In doing this we are effectively converting the 2MTF data from log-distance ratios to peculiar velocities to match the original Gaussian theory for the covariance matrix (Eq.~\ref{eq:cov2}), as opposed to changing the variable used in the theory from peculiar velocities to log-distance ratios (Eq.~\ref{eq:magfluccov}) to match the 2MTF data.

\begin{figure}
\centering
\includegraphics[width=0.5\textwidth]{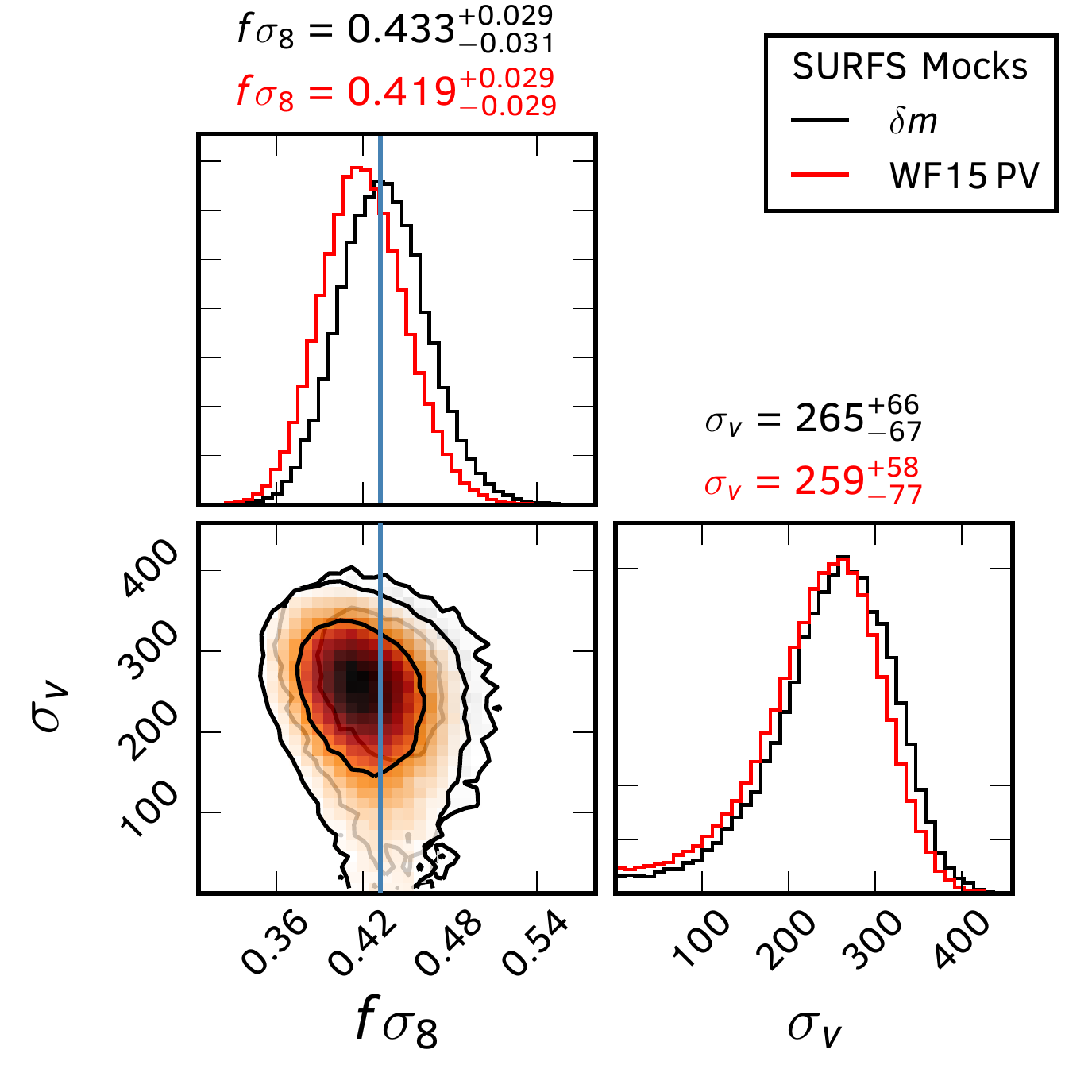} \\
  \caption{Likelihood contours and 1D marginalised histograms for fits to the SURFS mocks using the $\delta m$ variable and the peculiar velocity estimated using the WF15 peculiar velocity estimator (grey and red contours/histograms respectively). In both cases the data is gridded using $L=20\mpcoh$ and fit with $k_{max}=0.1\mpcoh$. The correct value for $f\sigma_{8}$ based on our fiducial cosmology is shown as the blue vertical line. Both methods return good results, with the WF15 estimator biased by only $0.013$ compared to an expected statistical error on the data of $0.082$. However we find that the $\delta m$ variable performs consistently better, being unbiased for the above case and closer to the true value for all $k_{max}$ tested.}
  \label{fig:SURFSfitsvel}
\end{figure}

Reproducing the fits using a grid size of $20\mpcoh$, we find that using the WF15 estimator or the $\delta m$ variable gives similar constraints, showing the same trend of increasing bias with increasing $k_{max}$. However we do find all the values of $f\sigma_{8}$ to be slightly low compared to the $\delta m$ variable. This shift is small but systemic, occurring even for $k_{max}=0.1\hompc$, which was previously unbiased. This case is shown in Fig.~\ref{fig:SURFSfitsvel}. Using the WF15 PV estimator we find a shift in the best-fit value (and expected statistical error in the data) of 0.013 (0.082) and 0.027 (0.072) for $k_{max}=0.1\hompc$ and $0.15\hompc$ respectively. This is compared to 0.001 (0.084) and 0.017 (0.075) for the same fits using the $\delta m$ variable. Hence we conclude that whilst the WF15 estimator does not return particularly biased results, its performance on the 2MTF data is not quite as good as using $\delta m$.

\begin{figure}
\centering
\includegraphics[width=0.5\textwidth]{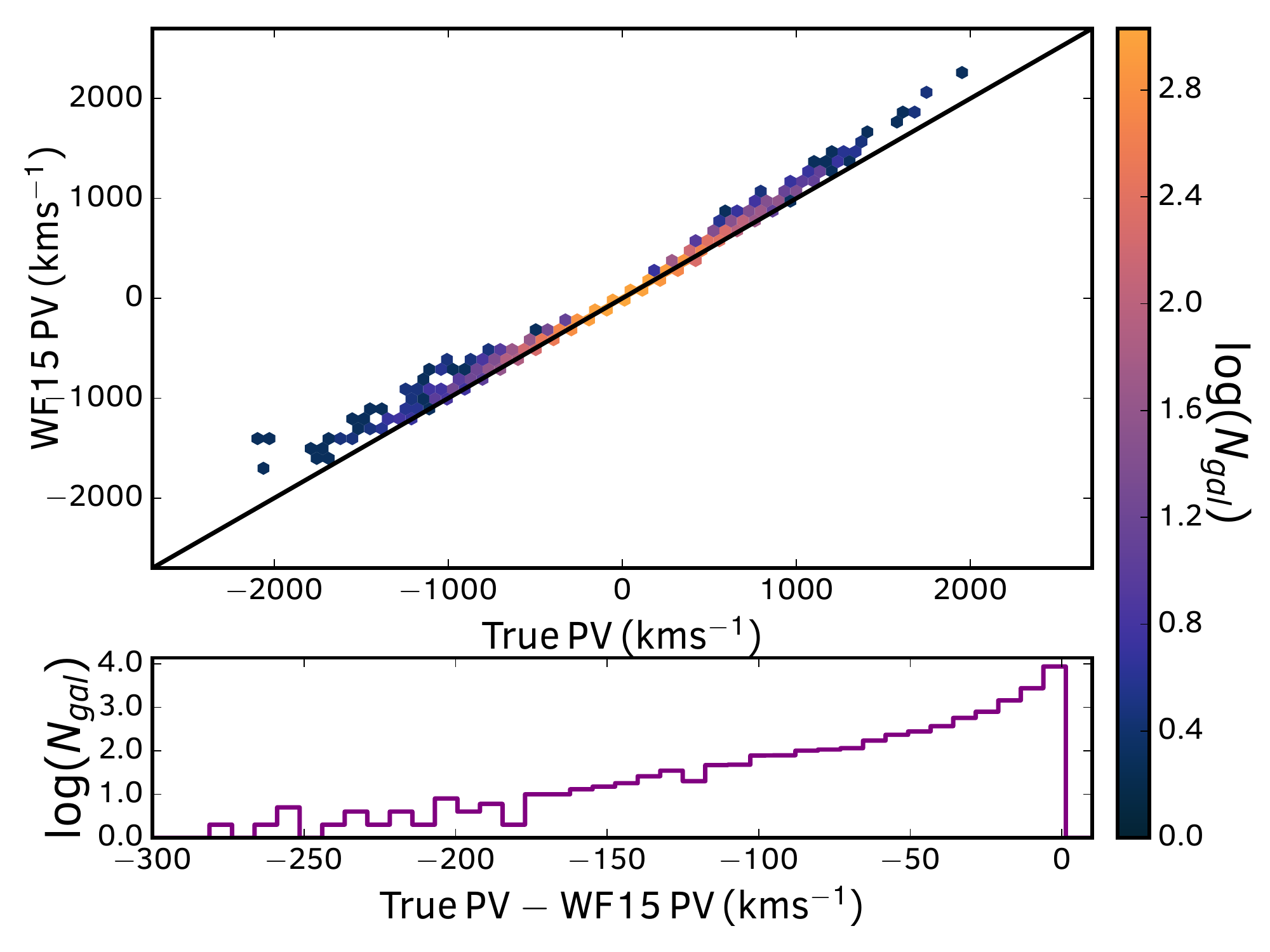} \\
  \caption{Plots comparing the true PVs measured from all of our mock catalogues against the PVs estimated using the WF15 estimator (Eq.~\ref{eq:wf15}) and the log-distance ratio of each galaxy in the mocks. The top panel shows the relationship between these two velocities, with the galaxies binned for clarity and the color of each point representing the number of galaxies in that bin. The solid black line is the 1:1 relationship we would expect to recover if the WF15 estimator returned the true PV. Instead we find some deviation from this line; large positive PVs are generally overestimated, whilst large negative PVs are underestimated. In the lower panel we plot a histogram of the difference between the true PV and the PV estimated with the WF15 estimator, which in the majority of cases are very small. There is no obvious link between the estimated velocities and a systematically low value of $f\sigma_{8}$; the WF15 estimator still recovers the width of the distribution of PVs well.}
  \label{fig:vels}
\end{figure}

We can use the mock catalogues to take a closer look at how the WF15 estimator behaves and if there is an obvious cause for this small systematic shift by comparing the known PVs of the galaxies in the mocks against those calculated using the log-distance ratio and the WF15 estimator. Fig.~\ref{fig:vels} plots the true PV of each mock galaxy against the estimated PV. We find that the WF15 estimator generally does very well, however it is not perfect, overestimating and underestimating large positive and negative peculiar velocities respectively. Unfortunately, how this translates to the small offset in $f\sigma_{8}$ seen in our fits is not clear. We expect the value of $f\sigma_{8}$ to be sensitive to the full distribution of velocities, such that underestimating the width of this distribution would return a low value of $f\sigma$. Looking at the mocks, whilst the mean velocity is offset slightly, the width of this distribution is still very well recovered and so this does not provide an explanation for the $f\sigma_{8}$ discrepancy.

To conclude, we find nothing in the behaviour of this estimator to suggest it should return a value of $f\sigma_{8}$ biased low. Whilst the WF15 estimator is sub-optimal compared to using $\delta m$ for measuring the velocity power spectrum in the 2MTF data, it is still generally unbiased and it is unclear whether other observables using the 2MTF measurements, such as the bulk flow, or velocity power spectrum measurements using other datasets, would find similar results. Hence the choice of variable for a given dataset remains one largely of convenience; using the WF15 estimator, particularly in bulk flow analyses, allows for a more intuitive understanding of the data than working in terms of magnitude fluctuations. These tests should be revisited for datasets with more constraining power to ensure that this estimator remains unbiased.

\subsection{Zero-point offsets} \label{sec:testssystematics}
Having decided on the variable and models we will use to fit the 2MTF data, we check whether our results are affected by systematics in the data and modelling. We fit the mocks with the same two methods identified above (gridded and ungridded with RSD modelling), but without marginalising over the effect of a velocity monopole and find negligible differences in the constraints. This corroborates the result of \cite{Howlett2017} and indicates that any systematic error in the velocity power spectrum associated with an offset in the zero-point is negligible compared to the measurement errors. Similar results were found by \cite{Johnson2014} in their fits to the 6dFGSv data.

\begin{table*}
\setlength{\extrarowheight}{3pt}
\caption{Constraints on the growth rate, $f\sigma_{8}$, non-linear velocity dispersion, $\sigma_{v}$ and non-linear RSD parameter $\sigma_{u}$ from fitting the velocity power spectrum in the 2MTF data for two different cosmologies, using the two different methods verified in Section~\ref{sec:tests}, and for the three photometric bands and the ``minimum error'' measurement separately.}
\centering
\begin{tabular}{lllccc} \hline
Cosmology & Method & 2MTF data & $f\sigma_{8}$ & $\sigma_{v} (\mathrm{km\,s^{-1}})$ & $\sigma_{u} (\mpcoh)$ \\ \hline \hline
\multirow{9}{*}{Fiducial (Planck-based)} & \multirow{4}{*}{Grid=$20\mpcoh$, $k_{max}=0.15\hompc$} & $K$-Band & $0.524^{+0.107}_{-0.086}$ & $338^{+122}_{-263}$ & - \\
 & & $H$-Band & $0.522^{+0.099}_{-0.091}$ & $325^{+168}_{-215}$ & - \\
 & & $J$-Band & $0.487^{+0.093}_{-0.080}$ & $<254$ & - \\
 & & ``Minimum Error'' & $0.491^{+0.090}_{-0.087}$ & $254^{+129}_{-225}$ & - \vspace{3pt} \\ 
 \cline{2-6} \vspace{-10pt} \\
 & \multirow{4}{*}{No Grid, $k_{max}=0.20\hompc$, RSD} & $K$-Band & $0.490^{+0.110}_{-0.061}$ & $186^{+28}_{-31}$ & $<7.96$ \\
 & & $H$-Band & $0.528^{+0.86}_{-0.076}$ & $213^{+24}_{-35}$ & $<6.87$  \\
 & & $J$-Band & $0.531^{+0.102}_{-0.071}$ & $192^{+25}_{-39}$ & $<7.56$  \\
 & & ``Minimum Error'' & $0.505^{+0.089}_{-0.079}$ & $196^{+25}_{-35}$ & $<8.49$ \vspace{3pt}\\ \hline 
\multirow{2}{*}{WMAP-based} & Grid=$20\mpcoh$, $k_{max}=0.15\hompc$ & ``Minimum Error'' & $0.468^{+0.089}_{-0.083}$ & $<342$ & -  \\ 
 & No Grid, $k_{max}=0.20\hompc$, RSD & ``Minimum Error'' & $0.473^{+0.094}_{-0.064}$ & $192^{+29}_{-31}$ & $<7.76$ \vspace{3pt} \\ \hline
\end{tabular}
\label{tab:results}
\end{table*}

\section{Results} \label{sec:results}

\begin{figure}
\centering
\includegraphics[width=0.5\textwidth]{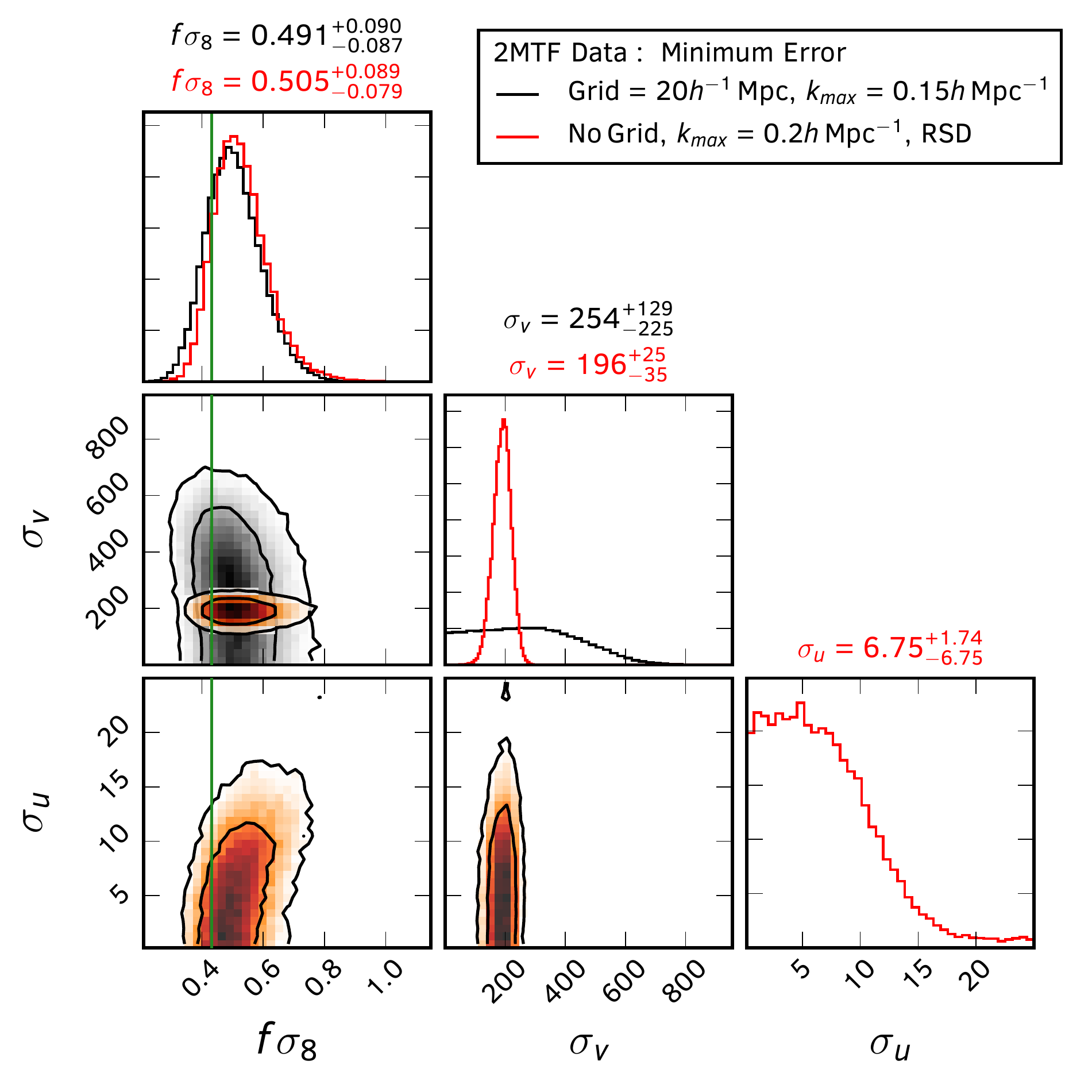} \\
  \caption{Likelihood and 1D marginalised constraints from fits to the 2MTF ``minimum error'' distances for our fiducial cosmology and using the methods tested throughout Section~\ref{sec:tests}. Our fits using the gridding (black) and non-linear RSD methods (red) are consistent, and in good agreement with the predictions of GR (vertical green line), though with a slight preference for higher growth rates.}
  \label{fig:2MTFresults}
\end{figure}

Having tested and finalised our fitting method in the previous section we now turn to fitting the 2MTF data. All our constraints are summarised in Table.~\ref{tab:results}. We plot the ``minimum error'' constraints for our fiducial cosmology and for both of our fitting methods in Fig.~\ref{fig:2MTFresults}. For our fiducial fitting methodology we ultimately obtain a measurement of $f\sigma_{8}(z=0) = 0.505^{+0.089}_{-0.079}$, a $\sim16\%$ measurement of the growth rate. A comparison of this result with other measurements of the growth rate is given in Section~\ref{sec:conclusions}.

In more detail, all our fits are performed using the magnitude fluctuation variable, $\delta m$ and marginalising over any velocity monopole. As with the mock results, we find negligible difference between the results using data with or without the correction or Malmquist bias, so only present results for the former here. As described in Section~\ref{sec:3bandscomb}, we use $4\sigma$-clipping, fit the $K$, $H$ and $J$-band distances separately, and also look at the case where we take the measurement with the smallest error for each galaxy. Our results for the 3 photometric bands and the ``minimum error'' measurements are all consistent. We also find consistent results using both of our fitting methods: gridding the data on scales of $20\mpcoh$ and fitting to $k_{max}=0.15\hompc$; and fitting the data without any gridding using $k_{max}=0.20\hompc$ and marginalising over the effects of non-linear RSD. Neither method produces consistently higher or lower values of $f\sigma_{8}$ in the data, indicating both of these methods are robust. When gridding the data, the constraints on the non-linear velocity dispersion are considerably weaker than when we use our non-linear RSD method as we are smoothing out information on scales where we could constrain this, but in all cases we find values consistent with those found in the mock catalogues and with the typical expectations of $150-300\mathrm{km\,s^{-1}}$. 

We find no strong preference for non-linear RSD damping in the 2MTF data. The maximum likelihood values for $\sigma_{u}$ are all consistent with zero, and the growth rate constraints are no weaker when we marginalise over this than for the gridding method. That is not to say that including this parameter is unnecessary; from our tests on the mocks we know neglecting to marginalise over it can bias results. Rather, the 2MTF data is too noisy on non-linear scales to constrain this parameter. Note that the mock constraints on $\sigma_{u}$ in Section~\ref{sec:tests} are from fitting all eight mocks simulaneously which gives significantly more constraining power than would be expected in the data, so the fact we constrain this parameter in Section~\ref{sec:tests} does not necessarily indicate we would expect to constrain it in the data.

In obtaining our constraints using the 2MTF data we have to assume some underlying fiducial cosmology, to both generate our model velocity power spectrum and convert each galaxies position into cartesian coordinates. To test the dependence of our results on the choice of fiducial cosmology we re-fit the data using a cosmology based on the results of WMAP \citep{Bennett2013}. This cosmology has the parameters $\Omega_{m}=0.273$, $\Omega_{b}=0.0456$, $H_{0}=70.5\mathrm{km\,s^{-1}\,Mpc^{-1}}$, $n_{s}=0.96$ and $\sigma_{8}=0.812$. The \textit{expected} value of the normalised growth rate, under the \textit{assumption} of GR, is $f\sigma_{8}=0.398$. Our measurements for this cosmological model are also shown in Table.~\ref{tab:results}, and will be compared to the prediction of GR in Section~\ref{sec:gamma}.

\begin{figure}
\centering
\includegraphics[width=0.5\textwidth]{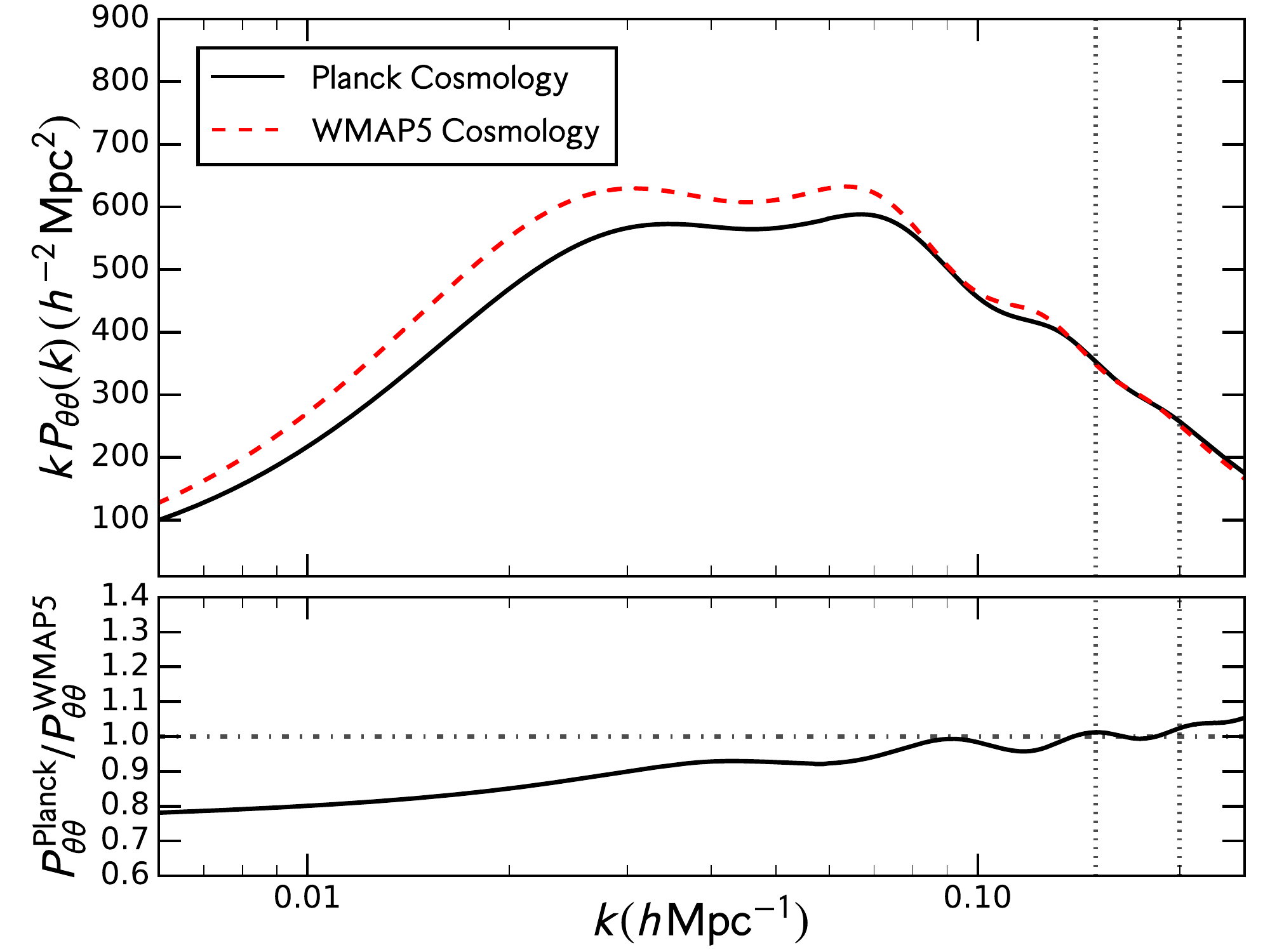} \\
  \caption{A plot of the velocity divergence power spectra (multiplied by $k$ to highlight the large scale differences) for the two cosmological models used in this work: our fiducial, Planck-based model (solid black line), and a WMAP-based cosmology (dashed red line). The ratio of the two power spectra is shown in the lower panel. The vertical dotted lines show $k=0.15\hompc$ and $0.2\hompc$, the smallest scales we fit against. The WMAP cosmology has more power on larger scales and less non-linear power, however the 2MTF data is mainly sensitive to linear scales where the difference between the cosmologies is approximately constant. Hence the increase in linear power in the WMAP cosmology can be counteracted with a lower value of $f\sigma_{8}$ to give similar velocity power spectra.}
  \label{fig:pkvel}
\end{figure}

When we fit the data using the WMAP cosmology we do find some evidence that the choice of cosmology impacts the $f\sigma_{8}$ constraints. However, this change can be understood by comparing the different velocity divergence power spectra for the two cosmologies in Fig.~\ref{fig:pkvel}, and by looking at the usual parameterisation of the growth rate $f=\Omega_{m}^{\gamma}$ \citep{Linder2007}. The larger value of $\Omega_{m}$ causes an decrease in the amplitude of the power spectrum on linear scales where we have the most constraining power, but the corresponding larger value of $f\sigma_{8}$ actually means that the resultant velocity power spectra are quite similar. Or, in other words, for a measured velocity power spectrum, such as that from the 2MTF data, changing the cosmology is effectively the same as changing the amplitude of the velocity divergence power spectrum, which can be compensated for by changing the growth rate, and so we measure a lower value of $f\sigma_{8}$ if we lower the value of $\Omega_{m}$. This is not generally true, as changing the cosmology also changes the shape of the velocity divergence power spectrum, but the weak constraining power of the 2MTF data on non-linear scales means that it is largely immune to this effect.  

Overall, because the change in $f\sigma_{8}$ when using the wrong cosmology can be partially explained by the change in $\Omega_{m}$, the resultant value of $\gamma$ and the consistency check of GR will not be biased as significantly as $f\sigma_{8}$. As the change in the growth rate is still $<0.5\sigma$, we hence find the 2MTF constraints to be robust to the choice of fiducial cosmology. In the next section we will use the $f\sigma_{8}$ constraints to measure $\gamma$ itself and show that these are consistent for both cosmologies. Nonetheless, this raises the point that other analyses, past or future, should carefully check the impact of assuming a fixed cosmological model on their measurements of the growth rate. Other datasets more sensitive to the actual shape of the velocity divergence power spectrum may suffer from this source of systematic error.

Based on the above considerations, and the fact that all our fits are ultimately consistent, we choose as our quoted constraint the fits using our fiducial cosmology with the non-linear RSD modelling and ``minimum error'' 2MTF measurements. In this case we arrive at our main result of $f\sigma_{8}(z=0) = 0.505^{+0.089}_{-0.079}$.

\subsection{Consistency with GR: $\gamma$ constraints} \label{sec:gamma}

Using our growth rate constraints we also perform a consistency check of GR by measuring the $\gamma$ parameter. We use the full likelihood for our ``minimum error'' fits using the non-linear RSD modelling without gridding and for the two different cosmologies in combination with the publicly available \textit{Planck}\footnote{For \textit{Planck} we use the base\_plikHM\_TTTEEE\_lowTEB\_lensing chain found at \url{https://wiki.cosmos.esa.int/planckpla/index.php/Cosmological_Parameters}} \citep{Planck2016} and WMAP9\footnote{For WMAP9 we use the $\Lambda$CDM MCMC chain found at \protect{\url{https://lambda.gsfc.nasa.gov/product/map/dr5/params/lcdm_wmap9.cfm}}} \citep{Hinshaw2013} likelihood chains. We use the method detailed in \cite{Howlett2015a} and at each likelihood evaluation we randomly sample from the CMB chain and randomly choose a value $0 < \gamma  \le 2$. The CMB chain provides values for $\Omega_{m}$ and $\sigma_{8}(z=0)$. However, the CMB is actually sensitive to the value $\sigma_{8}(z*)$ where $z*$ is the redshift of recombination, and the value of $\sigma_{8}(z=0)$ in the CMB chain is derived from this \textit{assuming GR}. Hence we scale each value of $\sigma_{8}(z=0)$ back to $z*$ using the linear growth factor for GR, then calculate the corresponding value of $\sigma^{\gamma}_{8}(z=0)$ under the new value of $\gamma$ using the correct growth factor. Even though we have fixed $\Omega_{m}$ for our 2MTF fits, the cosmologies used are very close to the maximum likelihood values for the chains we use. This means that we have simply neglected any additional information that the 2MTF data might give us about $\Omega_{m}$, which is perfectly valid as the combined likelihood for the background cosmology will be completely dominated by the CMB constraints anyway.

We also consider the $\gamma$ constraints when we combine our $f\sigma_{8}$ measurement with the results from the tomographic weak lensing analysis of the Kilo Degree Survey (KiDS; \citealt{Hildebrandt2017}). The constraints on $\Omega_{m}$ and $\sigma_{8}$ (in particular through the combination $\sigma_{8}\sqrt{\Omega_{m}/0.3}$) are in $\sim2.3\sigma$ tension with the results from \cite{Planck2016}, hence it is interesting to study the consistency of their results and ours with GR. We obtain constraints on $\gamma$ using the same method as for the \textit{Planck} and WMAP chains\footnote{The KiDS chains can be found at \url{http://kids.strw.leidenuniv.nl/cosmicshear2016.php}}, however as the KiDS results are sensitive to the present day value of $\sigma_{8}$ (unlike the CMB where $\sigma_{8}$ is extrapolated to the present day assuming GR) we do not correct the value of $\sigma_{8}$ to account for the different values of $\gamma$. One important caveat to this is that the KiDS results provide much weaker constraints on $\Omega_{m}$ than \textit{Planck} or WMAP9, so the effects of using a fixed value of $\Omega_{m}$ in our fits to the 2MTF data may be more important. We choose to use the 2MTF constraints given the WMAP-based input cosmology  (i.e., the last line in Table.~\ref{tab:results}) as this is closer to the maximum likelihood cosmology for the KiDS data and should reduce any potential biases, but note that a rigorous combination of these two datasets should allow for the fact that the 2MTF data may provide additional information on $\Omega_{m}$ beyond that available with just KiDS.

\begin{figure}
\centering
\includegraphics[width=0.5\textwidth]{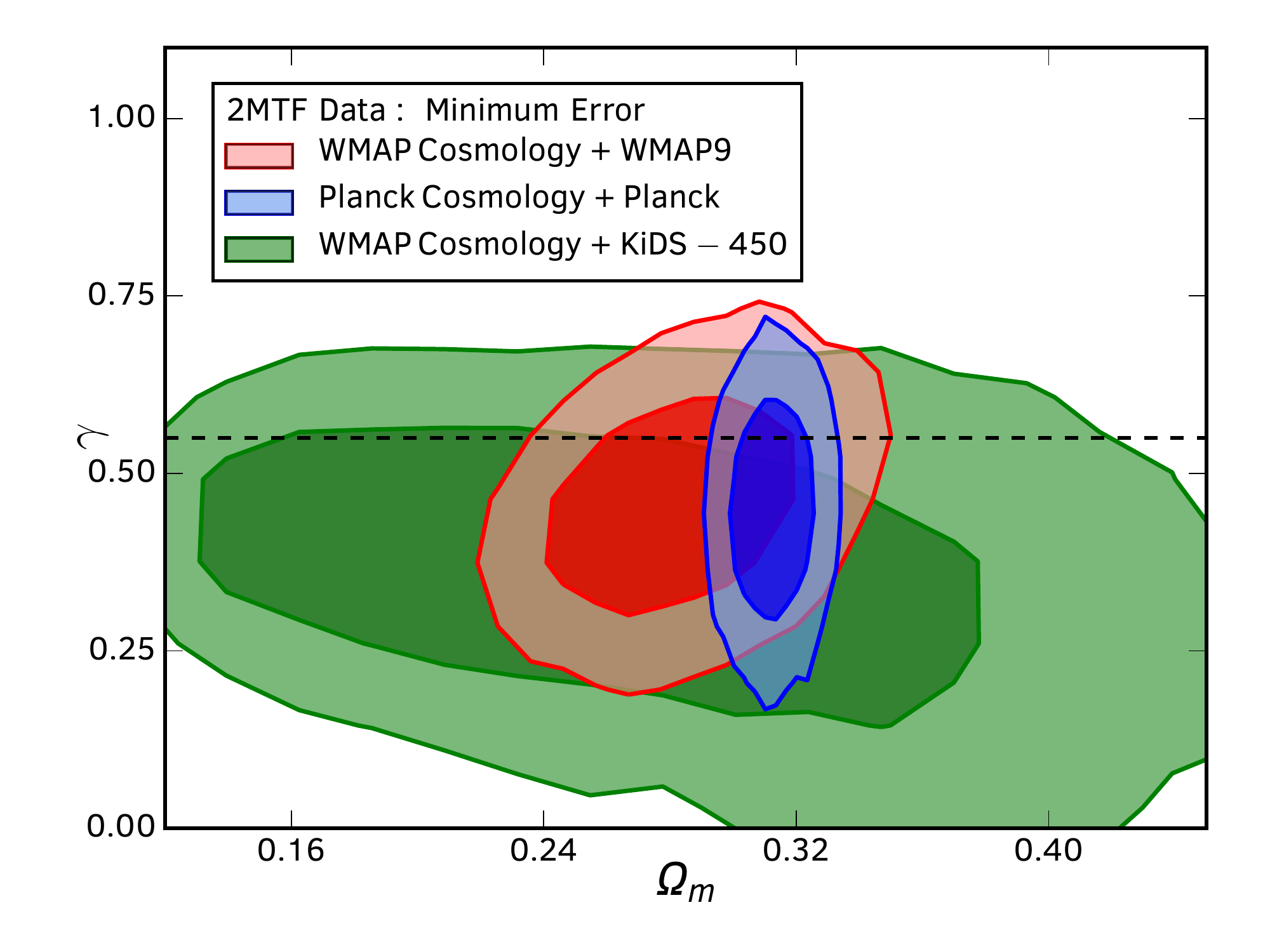} \\
  \caption{$1$ and $2\sigma$ likelihood contours for $\Omega_{m}$ and $\gamma$ from the combination of 2MTF ``minimum error'' growth rate measurements with \textit{Planck} (blue), WMAP9 (red) and KiDS (green) data. In all cases we use the 2MTF measurements without gridding and including non-linear RSD modelling. For combining with \textit{Planck} we use the 2MTF result for a Planck-based fiducial cosmology (i.e., the eighth row in Table.~\ref{tab:results}), whereas for combining with WMAP9 or KiDS we use the 2MTF results for a WMAP-based fiducial cosmology (the last row in Table.~\ref{tab:results}), as this is closer the the maximum likelihood cosmology for these two datasets. The dashed horizontal line is the prediction from GR. All cases are self-consistent and in good agreement with GR, though with a preference for lower $\gamma$.}
  \label{fig:2MTFgamma}
\end{figure}

The 2D likelihood in the $\Omega_{m}$-$\gamma$ plane is shown in Fig.~\ref{fig:2MTFgamma}. We find marginalised values of $\gamma = 0.45^{+0.10}_{-0.11}$ for both the WMAP9 and \textit{Planck} cosmologies and $\gamma=0.38^{+0.12}_{-0.15}$ for the 2MTF results assuming a WMAP-based input cosmology combined with the results of KiDS. That the combined 2MTF and CMB results using the two cosmologies are so similar proves that although the use of different cosmologies changes the values of $f\sigma_{8}$ measured from the data, this is mainly due to the change in $\Omega_{m}$ and is hence accounted for when we calculate $\gamma$. The combination of our 2MTF result with the KiDS data generally prefers lower $\Omega_{m}$ and $\gamma$ than the combination with CMB data, although the error bars are large. In general, we conclude that all three values are consistent with the predictions of GR ($\gamma\approx 0.55)$ within $\sim1.5\sigma$, although with a preference for lower $\gamma$.

\subsection{Scale-dependent constraints}

\begin{figure*}
\centering
\begin{tabular}[b]{l}
\includegraphics[width=0.57\textwidth]{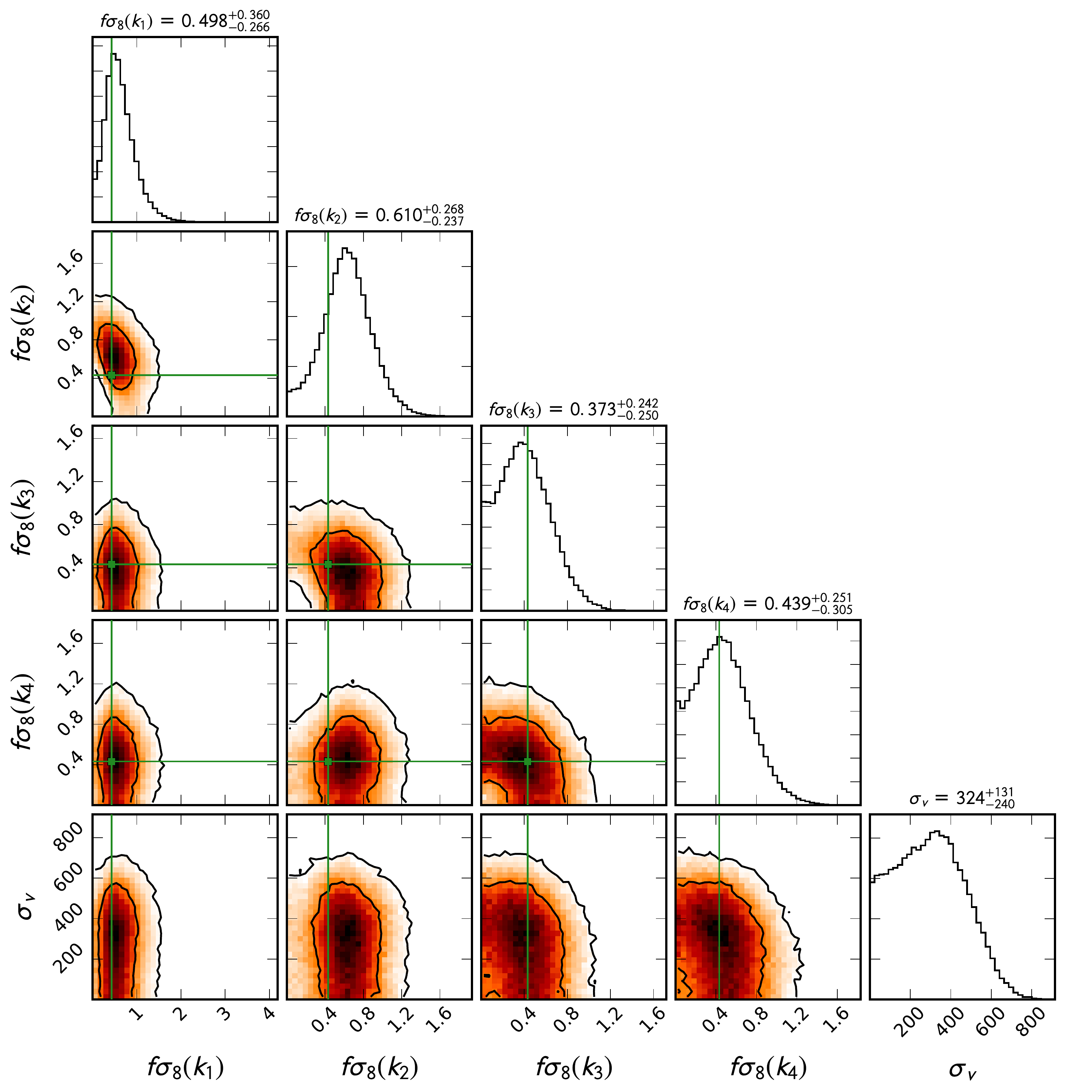}
\end{tabular}
\begin{tabular}[b]{r}
\includegraphics[width=0.37\textwidth]{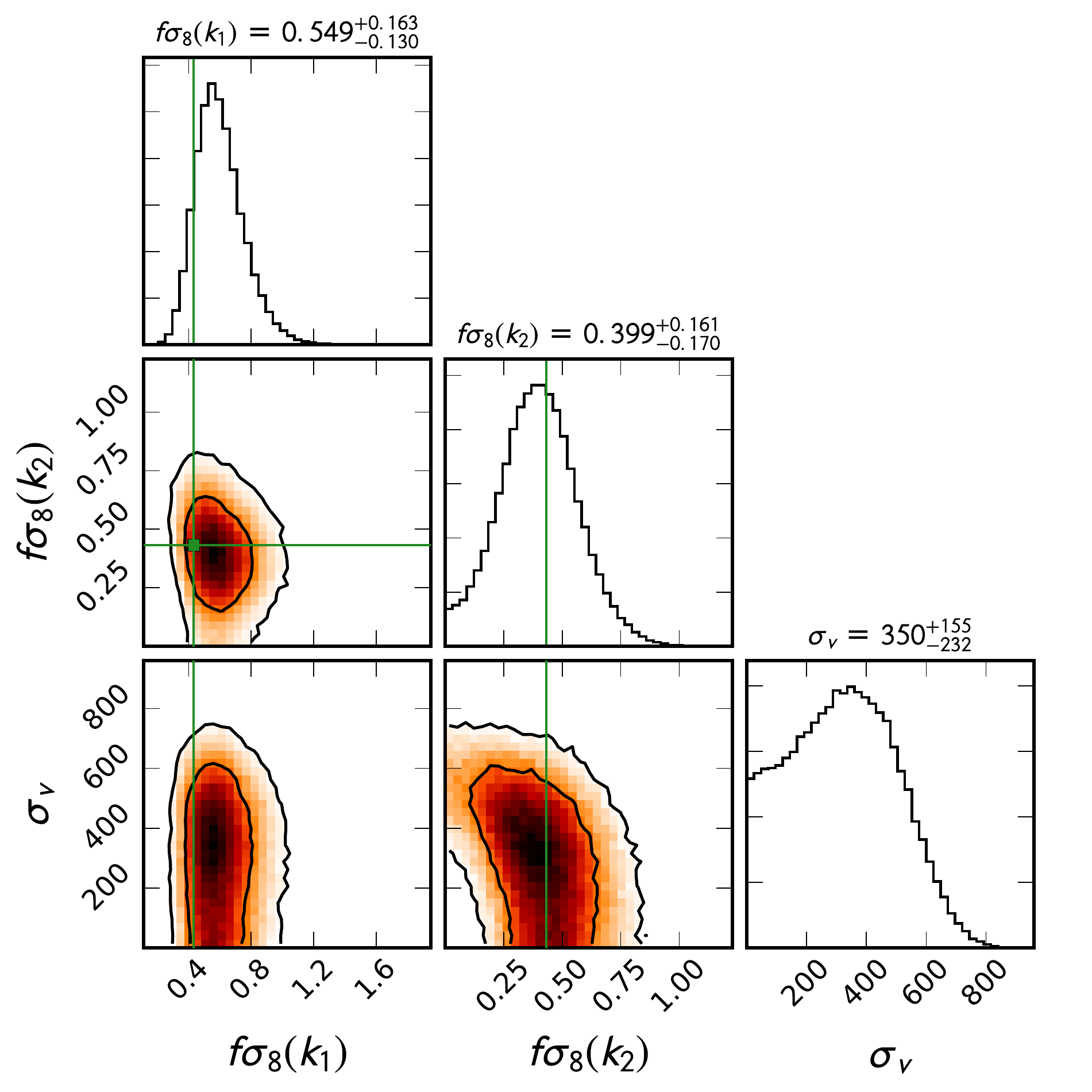}
\end{tabular}
  \caption{Likelihood contours and 1D marginalised histograms from \textit{scale-dependent} fits to the 2MTF ``minimum error'' distances for our fiducial cosmological model. We fit the velocity power spectrum in four $k$-ranges: $k_{1}=0.007-0.025\hompc$, $k_{2}=0.025-0.055\hompc$, $k_{3}=0.055-0.105\hompc$ and $k_{4}=0.105-0.150\hompc$ in the left plot and two $k$-ranges: $k_{1}=0.007-0.055\hompc$ and $k_{2}=0.055-0.150\hompc$ in the right plot. In all cases we find results that are consistent with the predictions on GR (green horizontal and vertical lines) although with a slight preference for larger growth rates in our $k_{2}$-bin and $k_{1}$-bin when using four and two bins respectively. These are consistent with statistical fluctuations and an excess of power is not seen on any larger scales when using four bins.}
  \label{fig:2MTFksdcontours}
\end{figure*}

On top of measuring a scale-free value for the growth rate, we can model the velocity power spectrum is discrete $k$-bins to obtain \textit{scale-dependent} constraints, as was done by both \cite{Macaulay2012} and \cite{Johnson2014}. GR predicts a scale-free growth rate, so any observed scale-dependence would point to modifications to gravity. Including non-linear RSD damping in our model will primarily affect the constraints on small-scales, with $\sigma_{u}$ being strongly degenerate with the growth rate in this regime. Hence, for our scale-dependent constraints we decide to use the gridding method tested and verified in Section~\ref{sec:tests} and which was already found to produce consistent measurements of the growth rate using the 2MTF data. 

\begin{figure}
\centering
\includegraphics[width=0.5\textwidth]{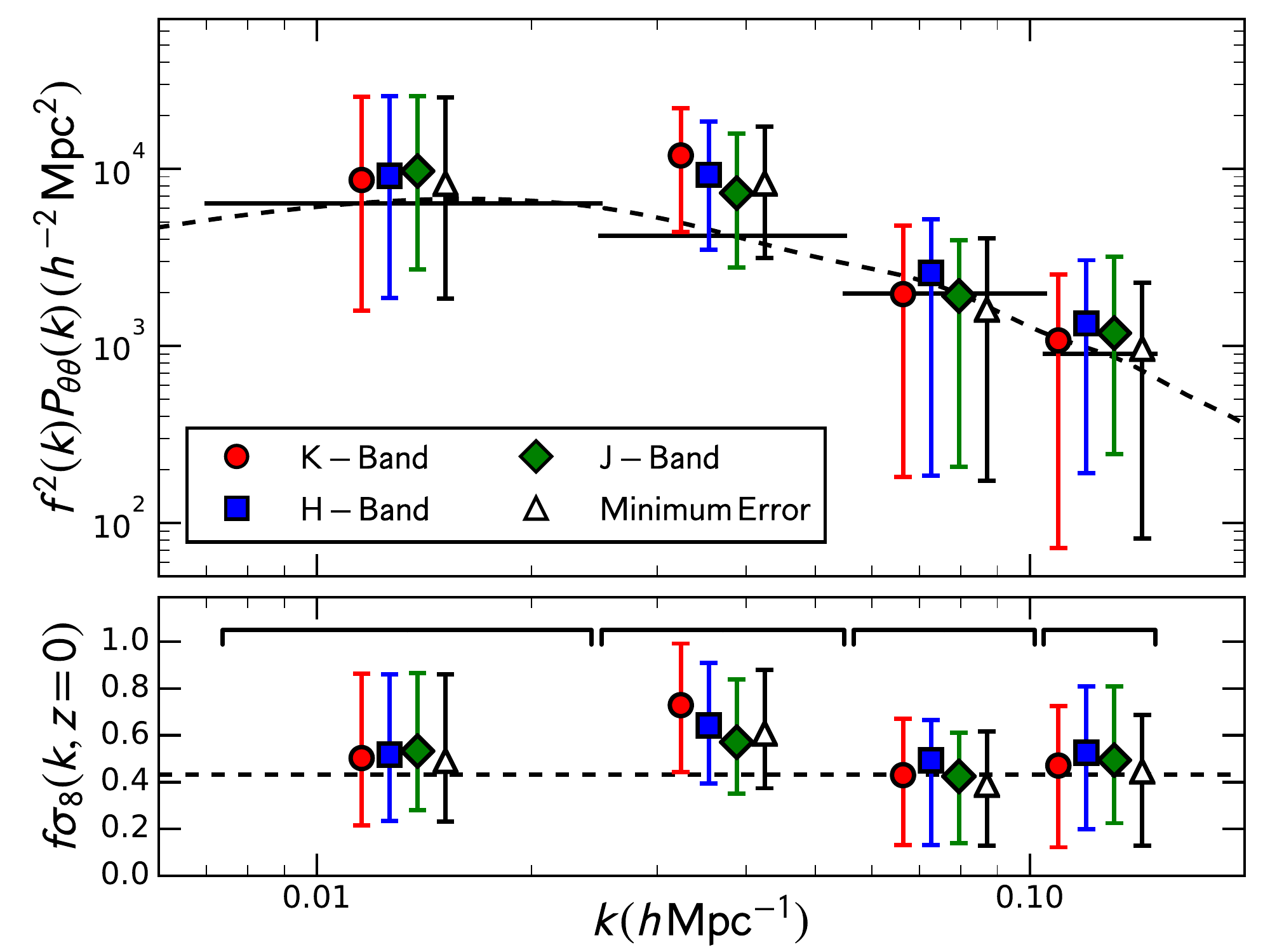} \\
  \caption{Best-fit and $1\sigma$ errors on the growth rate in four $k$-bins and for the 3 photometric bands and the ``minimum error'' distances. In the top panel we plot the growth rate constraints multiplied by our fiducial velocity divergence power spectrum. The expectation assuming GR is plotted as the dashed line. Our fits are sensitive to the power and growth rate averaged across each bin. The corresponding GR expectation is shown as a solid horizontal line. The position of the points relative to this line is arbitrary. In the lower panel we plot the constraints on the growth rate itself, alongside the scale-independent prediction from GR (dashed line). In all cases we find results statistically consistent with GR.}
  \label{fig:2MTFksdbestfit}
\end{figure}

Our scale dependent constraints are obtained by fitting the velocity power spectrum in four $k$-ranges. We also look at the case where we use only two bins. The likelihood contours and 1D marginalised histograms for fits to the ``minimum error'' 2MTF data are shown in Fig~\ref{fig:2MTFksdcontours}. The exact $k$-bins and $f\sigma_{8}(k,z=0)$ constraints are given in Table~\ref{tab:fsigma8sd}. The constraints are generally consistent with the expected value for our fiducial cosmology, with some small preference for a larger than expected growth rate in our second $k$-bin when using four. We find no evidence for an excess of power on scales larger than this, and our results using two bins are also fully consistent with GR. The maximum likelihood and $1\sigma$ errors for the four-bin fit, using the ``minimum error" 2MTF data and for the 3 individual photometric bands, are also compared to our fiducial cosmology in Fig.~\ref{fig:2MTFksdbestfit}, where we plot the velocity divergence power spectrum for our fiducial cosmology multiplied by the expected growth rate. We find the preference for larger power in the $k_{2}$-bin in all of our photometric bands, but this remains consistent with statistical fluctuations, and again there is no evidence in any of the bands for such a preference on larger scales.

\begin{table}
\setlength{\extrarowheight}{3pt}
\caption{Scale dependent constraints on the growth rate, $f\sigma_{8}(k)$ at $z=0$ from fitting the velocity power spectrum of the ``minimum error'' 2MTF data in different $k$-bins. Columns give the upper and lower limits of each bin and the corresponding $f\sigma_{8}$ measurement. The upper and lower segments of the table give the case for two and four $k$-bins respectively.}
\centering
\begin{tabular}{lc} \hline
$k$-range ($\hompc$) & $f\sigma_{8}(k,z=0)$ \\ \hline \hline
$0.007-0.055$ & $0.549^{+0.163}_{-0.130}$ \\
$0.055-0.150$ & $0.399^{+0.161}_{-0.170}$ \vspace{3pt} \\ \hline
$0.007-0.025$ & $0.498^{+0.360}_{-0.266}$ \\
$0.025-0.055$ & $0.610^{+0.268}_{-0.237}$ \\ 
$0.055-0.105$ & $0.373^{+0.242}_{-0.250}$ \\
$0.105-0.150$ & $0.439^{+0.251}_{-0.305}$ \vspace{3pt} \\ 
 \hline
\end{tabular}
\label{tab:fsigma8sd}
\end{table}

\section{Discussion and Conclusion} \label{sec:conclusions}
In this work we have presented measurements of the velocity power spectrum using a set of 2062 measured peculiar velocities in the completed 2MASS Tully Fisher Survey. We have performed a rigorous test of our fitting methodology using a set of eight realistic mock 2MTF surveys. We identify the regimes in which our method returns unbiased fits, and introduce a greater degree of flexibility in our modelling than was used in previous studies. This is done by accounting for and marginalising over the effects of non-linear Redshift-Space Distortions. Our method is able to produce unbiased fits to smaller scales than is possible otherwise and than was used by previous studies, and without the need for gridding or smoothing the data.

We test two different Gaussian distributed variables that can be used to estimate the velocity power spectrum: magnitude fluctuations, $\delta_{m}$, which was the variable adopted in \cite{Johnson2014}, and which we also find produces reliable fits to the 2MTF data; and the peculiar velocity estimator of \cite{Watkins2015}. We find that this latter estimator is also generally unbiased but does seem to slightly underestimate the value of $f\sigma_{8}$ found in the mocks regardless of the scales fit against. Comparing the true and estimated peculiar velocities in the mocks reveals no obvious cause for this and this discrepancy is not statistically significant. We decide to use the $\delta m$ variable in our fits as we find it to be superior for the 2MTF data. Whether this estimator is nonetheless effective, or even better, for estimating the bulk flow in the 2MTF data or for use on other datasets remains an open question.

We verify that our fits are robust to the effects of a change in cosmology or a velocity monopole. That said, this may only be true because the constraining power of 2MTF is sufficiently weak. For future PV surveys, such as Taipan \citep{daCunha2017} or WALLABY \citep{Johnston2008}, with much greater constraining power we may have to marginalise over the effects of different cosmological models on the growth rate constraints.

Overall, we find best-fit \textit{scale-dependent} constraints on the growth rate of structure at redshift zero that are consistent with a scale-independent growth rate and the prediction of GR, when using both two and four $k$-bins. \textit{Assuming} scale-independence we find a value $f\sigma_{8}(z=0) = 0.505^{+0.089}_{-0.079}$, also consistent with the predictions of GR. This is a $\sim16\%$ measurement of the growth rate, comparable to the constraints using the 6dFGSv sample of $\sim8,800$ galaxies (\citealt{Johnson2014}; $\sim 15\%$) and the constraints using the 6dFGRS (\citealt{Beutler2012}; $\sim 13\%$), which contains over $100,000$ galaxies. The fact that we obtain comparable results using a smaller number of galaxies is a result of the higher number density of local objects and better distance measurement compared to 6dFGSv, and the fact that PV surveys are independent of the effects of galaxy bias. Combining our growth rate measurements with CMB data from \textit{Planck} or WMAP9 we find $\gamma = 0.45^{+0.10}_{-0.11}$, a $\sim25\%$ measurement, consistent with GR. Combining with weak lensing measurements from KiDS we find, $\gamma=0.38^{+0.12}_{-0.15}$, which is consistent with GR at the level of $\sim 1.5\sigma$.

The fact that our constraints on $\gamma$ are only a factor of two larger than state-of-the-art constraints combining a number of Large Scale Structure, CMB and Type Ia supernovae measurements \citep{Mueller2016} highlights the strong tests of gravity that can be made using PV surveys, both because of their independence from galaxy bias and their low redshift. For future surveys containing larger numbers of both redshifts \textit{and} velocities, such as that planned with WALLABY \citep{Duffy2012,Koribalski2012}, these properties will enable growth rate measurements comparable to, and even surpassing, those that can be made using traditional Large Scale Structure surveys \citep{Howlett2017}.

\section*{Acknowledgements}
We thank Chris Power for providing access to the SURFS simulation. An earlier version of this work also relied heavily on the GiggleZ cosmological simulation kindly provided to us Greg Poole \citep{Poole2015}.

We also thank Chris Blake, Andrew Johnson and Caitlin Adams for their useful and insightful correspondence over the course of this work.

This research was conducted by the Australian Research Council Centre of Excellence for All-sky Astrophysics (CAASTRO), through project number CE110001020.

This work used data from the Robert C. Byrd Green Bank Radio Telescope obtained through observing projects GBT06A-027, GBT06B-021, GBT06C-049, GBT08B-003: ``Mapping Mass in the Nearby Universe with 2MASS'', PI Karen L. Masters.

Numerical computations were done on the Pleiades HPC cluster at the International Centre for Radio Astronomy Research at the University of Western Australia, and the Sciama High Performance Compute (HPC) cluster which is supported by the ICG, SEPNet and the University of Portsmouth.

This research has made use of NASA's Astrophysics Data System Bibliographic Services and the \texttt{astro-ph} pre-print archive at \url{https://arxiv.org/}. All plots in this paper were made using the {\sc matplotlib} plotting library \citep{Hunter2007}.

\bibliography{/Volumes/Work/ICRAR/LaTeX/massive.bib}{}

\begin{thebibliography}{}
\makeatletter
\relax
\def\mn@urlcharsother{\let\do\@makeother \do\$\do\&\do\#\do\^\do\_\do\%\do\~}
\def\mn@doi{\begingroup\mn@urlcharsother \@ifnextchar [ {\mn@doi@}
  {\mn@doi@[]}}
\def\mn@doi@[#1]#2{\def\@tempa{#1}\ifx\@tempa\@empty \href
  {http://dx.doi.org/#2} {doi:#2}\else \href {http://dx.doi.org/#2} {#1}\fi
  \endgroup}
\def\mn@eprint#1#2{\mn@eprint@#1:#2::\@nil}
\def\mn@eprint@arXiv#1{\href {http://arxiv.org/abs/#1} {{\tt arXiv:#1}}}
\def\mn@eprint@dblp#1{\href {http://dblp.uni-trier.de/rec/bibtex/#1.xml}
  {dblp:#1}}
\def\mn@eprint@#1:#2:#3:#4\@nil{\def\@tempa {#1}\def\@tempb {#2}\def\@tempc
  {#3}\ifx \@tempc \@empty \let \@tempc \@tempb \let \@tempb \@tempa \fi \ifx
  \@tempb \@empty \def\@tempb {arXiv}\fi \@ifundefined
  {mn@eprint@\@tempb}{\@tempb:\@tempc}{\expandafter \expandafter \csname
  mn@eprint@\@tempb\endcsname \expandafter{\@tempc}}}

\bibitem[\protect\citeauthoryear{{Abate}, {Bridle}, {Teodoro}, {Warren}  \&
  {Hendry}}{{Abate} et~al.}{2008}]{Abate2008}
{Abate} A.,  {Bridle} S.,  {Teodoro} L.~F.~A.,  {Warren} M.~S.,   {Hendry} M.,
  2008, \mn@doi [\mnras] {10.1111/j.1365-2966.2008.13637.x}, \href
  {http://adsabs.harvard.edu/abs/2008MNRAS.389.1739A} {389, 1739}

\bibitem[\protect\citeauthoryear{{Agarwal}, {Abdalla}, {Feldman}, {Lahav}  \&
  {Thomas}}{{Agarwal} et~al.}{2014}]{Agarwal2014}
{Agarwal} S.,  {Abdalla} F.~B.,  {Feldman} H.~A.,  {Lahav} O.,   {Thomas}
  S.~A.,  2014, \mn@doi [\mnras] {10.1093/mnras/stu090}, \href
  {http://adsabs.harvard.edu/abs/2014MNRAS.439.2102A} {439, 2102}

\bibitem[\protect\citeauthoryear{{Alam} et~al.,}{{Alam}
  et~al.}{2016}]{Alam2016}
{Alam} S.,  et~al., 2016, preprint, \href
  {http://adsabs.harvard.edu/abs/2016arXiv160703155A} {} (\mn@eprint {arXiv}
  {1607.03155})

\bibitem[\protect\citeauthoryear{{Azzalini} \& {Capitanio}}{{Azzalini} \&
  {Capitanio}}{2009}]{Azzalini2009}
{Azzalini} A.,  {Capitanio} A.,  2009, preprint, \href
  {http://adsabs.harvard.edu/abs/2009arXiv0911.2093A} {} (\mn@eprint {arXiv}
  {0911.2093})

\bibitem[\protect\citeauthoryear{{Bennett} et~al.,}{{Bennett}
  et~al.}{2013}]{Bennett2013}
{Bennett} C.~L.,  et~al., 2013, \mn@doi [\apjs] {10.1088/0067-0049/208/2/20},
  \href {http://adsabs.harvard.edu/abs/2013ApJS..208...20B} {208, 20}

\bibitem[\protect\citeauthoryear{{Beutler} et~al.,}{{Beutler}
  et~al.}{2012}]{Beutler2012}
{Beutler} F.,  et~al., 2012, \mn@doi [\mnras]
  {10.1111/j.1365-2966.2012.21136.x}, \href
  {http://adsabs.harvard.edu/abs/2012MNRAS.423.3430B} {423, 3430}

\bibitem[\protect\citeauthoryear{{Blake} et~al.,}{{Blake}
  et~al.}{2011}]{Blake2011}
{Blake} C.,  et~al., 2011, \mn@doi [\mnras] {10.1111/j.1365-2966.2011.18903.x},
  \href {http://adsabs.harvard.edu/abs/2011MNRAS.415.2876B} {415, 2876}

\bibitem[\protect\citeauthoryear{{Branchini} et~al.,}{{Branchini}
  et~al.}{1999}]{Branchini1999}
{Branchini} E.,  et~al., 1999, \mn@doi [\mnras]
  {10.1046/j.1365-8711.1999.02514.x}, \href
  {http://adsabs.harvard.edu/abs/1999MNRAS.308....1B} {308, 1}

\bibitem[\protect\citeauthoryear{{Branchini}, {Davis}  \& {Nusser}}{{Branchini}
  et~al.}{2012}]{Branchini2012}
{Branchini} E.,  {Davis} M.,   {Nusser} A.,  2012, \mn@doi [\mnras]
  {10.1111/j.1365-2966.2012.21210.x}, \href
  {http://adsabs.harvard.edu/abs/2012MNRAS.424..472B} {424, 472}

\bibitem[\protect\citeauthoryear{{Bridle}, {Crittenden}, {Melchiorri},
  {Hobson}, {Kneissl}  \& {Lasenby}}{{Bridle} et~al.}{2002}]{Bridle2002}
{Bridle} S.~L.,  {Crittenden} R.,  {Melchiorri} A.,  {Hobson} M.~P.,  {Kneissl}
  R.,   {Lasenby} A.~N.,  2002, \mn@doi [\mnras]
  {10.1046/j.1365-8711.2002.05709.x}, \href
  {http://adsabs.harvard.edu/abs/2002MNRAS.335.1193B} {335, 1193}

\bibitem[\protect\citeauthoryear{{Carlson}, {White}  \&
  {Padmanabhan}}{{Carlson} et~al.}{2009}]{Carlson2009}
{Carlson} J.,  {White} M.,   {Padmanabhan} N.,  2009, \mn@doi [\prd]
  {10.1103/PhysRevD.80.043531}, \href
  {http://adsabs.harvard.edu/abs/2009PhRvD..80d3531C} {80, 043531}

\bibitem[\protect\citeauthoryear{{Carrick}, {Turnbull}, {Lavaux}  \&
  {Hudson}}{{Carrick} et~al.}{2015}]{Carrick2015}
{Carrick} J.,  {Turnbull} S.~J.,  {Lavaux} G.,   {Hudson} M.~J.,  2015, \mn@doi
  [\mnras] {10.1093/mnras/stv547}, \href
  {http://adsabs.harvard.edu/abs/2015MNRAS.450..317C} {450, 317}

\bibitem[\protect\citeauthoryear{{Cole} \& {Kaiser}}{{Cole} \&
  {Kaiser}}{1989}]{Cole1989}
{Cole} S.,  {Kaiser} N.,  1989, \mn@doi [\mnras] {10.1093/mnras/237.4.1127},
  \href {http://adsabs.harvard.edu/abs/1989MNRAS.237.1127C} {237, 1127}

\bibitem[\protect\citeauthoryear{{Coles} \& {Lucchin}}{{Coles} \&
  {Lucchin}}{1995}]{ColesLucchin}
{Coles} P.,  {Lucchin} F.,  1995, {Cosmology. The origin and evolution of
  cosmic structure}

\bibitem[\protect\citeauthoryear{{Conroy}, {Wechsler}  \& {Kravtsov}}{{Conroy}
  et~al.}{2006}]{Conroy2006}
{Conroy} C.,  {Wechsler} R.~H.,   {Kravtsov} A.~V.,  2006, \mn@doi [\apj]
  {10.1086/503602}, \href {http://adsabs.harvard.edu/abs/2006ApJ...647..201C}
  {647, 201}

\bibitem[\protect\citeauthoryear{{Crocce} \& {Scoccimarro}}{{Crocce} \&
  {Scoccimarro}}{2006a}]{Crocce2006a}
{Crocce} M.,  {Scoccimarro} R.,  2006a, \mn@doi [\prd]
  {10.1103/PhysRevD.73.063519}, \href
  {http://adsabs.harvard.edu/abs/2006PhRvD..73f3519C} {73, 063519}

\bibitem[\protect\citeauthoryear{{Crocce} \& {Scoccimarro}}{{Crocce} \&
  {Scoccimarro}}{2006b}]{Crocce2006b}
{Crocce} M.,  {Scoccimarro} R.,  2006b, \mn@doi [\prd]
  {10.1103/PhysRevD.73.063520}, \href
  {http://adsabs.harvard.edu/abs/2006PhRvD..73f3520C} {73, 063520}

\bibitem[\protect\citeauthoryear{{Crocce} \& {Scoccimarro}}{{Crocce} \&
  {Scoccimarro}}{2008}]{Crocce2008}
{Crocce} M.,  {Scoccimarro} R.,  2008, \mn@doi [\prd]
  {10.1103/PhysRevD.77.023533}, \href
  {http://adsabs.harvard.edu/abs/2008PhRvD..77b3533C} {77, 023533}

\bibitem[\protect\citeauthoryear{{Davis} \& {Scrimgeour}}{{Davis} \&
  {Scrimgeour}}{2014}]{DavisT2014}
{Davis} T.~M.,  {Scrimgeour} M.~I.,  2014, \mn@doi [\mnras]
  {10.1093/mnras/stu920}, \href
  {http://adsabs.harvard.edu/abs/2014MNRAS.442.1117D} {442, 1117}

\bibitem[\protect\citeauthoryear{{Davis}, {Efstathiou}, {Frenk}  \&
  {White}}{{Davis} et~al.}{1985}]{DavisM1985}
{Davis} M.,  {Efstathiou} G.,  {Frenk} C.~S.,   {White} S.~D.~M.,  1985,
  \mn@doi [\apj] {10.1086/163168}, \href
  {http://adsabs.harvard.edu/abs/1985ApJ...292..371D} {292, 371}

\bibitem[\protect\citeauthoryear{{Davis}, {Nusser}  \& {Willick}}{{Davis}
  et~al.}{1996}]{DavisM1996}
{Davis} M.,  {Nusser} A.,   {Willick} J.~A.,  1996, \mn@doi [\apj]
  {10.1086/178124}, \href {http://adsabs.harvard.edu/abs/1996ApJ...473...22D}
  {473, 22}

\bibitem[\protect\citeauthoryear{{Davis}, {Nusser}, {Masters}, {Springob},
  {Huchra}  \& {Lemson}}{{Davis} et~al.}{2011a}]{DavisM2011}
{Davis} M.,  {Nusser} A.,  {Masters} K.~L.,  {Springob} C.,  {Huchra} J.~P.,
  {Lemson} G.,  2011a, \mn@doi [\mnras] {10.1111/j.1365-2966.2011.18362.x},
  \href {http://adsabs.harvard.edu/abs/2011MNRAS.413.2906D} {413, 2906}

\bibitem[\protect\citeauthoryear{{Davis} et~al.,}{{Davis}
  et~al.}{2011b}]{DavisT2011}
{Davis} T.~M.,  et~al., 2011b, \mn@doi [\apj] {10.1088/0004-637X/741/1/67},
  \href {http://adsabs.harvard.edu/abs/2011ApJ...741...67D} {741, 67}

\bibitem[\protect\citeauthoryear{{Desjacques} \& {Sheth}}{{Desjacques} \&
  {Sheth}}{2010}]{Desjacques2010}
{Desjacques} V.,  {Sheth} R.~K.,  2010, \mn@doi [\prd]
  {10.1103/PhysRevD.81.023526}, \href
  {http://adsabs.harvard.edu/abs/2010PhRvD..81b3526D} {81, 023526}

\bibitem[\protect\citeauthoryear{{Djorgovski} \& {Davis}}{{Djorgovski} \&
  {Davis}}{1987}]{Djorgovski1987}
{Djorgovski} S.,  {Davis} M.,  1987, \mn@doi [\apj] {10.1086/164948}, \href
  {http://adsabs.harvard.edu/abs/1987ApJ...313...59D} {313, 59}

\bibitem[\protect\citeauthoryear{{Dressler}, {Lynden-Bell}, {Burstein},
  {Davies}, {Faber}, {Terlevich}  \& {Wegner}}{{Dressler}
  et~al.}{1987}]{Dressler1987}
{Dressler} A.,  {Lynden-Bell} D.,  {Burstein} D.,  {Davies} R.~L.,  {Faber}
  S.~M.,  {Terlevich} R.,   {Wegner} G.,  1987, \mn@doi [\apj]
  {10.1086/164947}, \href {http://adsabs.harvard.edu/abs/1987ApJ...313...42D}
  {313, 42}

\bibitem[\protect\citeauthoryear{{Duffy}, {Meyer}, {Staveley-Smith}, {Bernyk},
  {Croton}, {Koribalski}, {Gerstmann}  \& {Westerlund}}{{Duffy}
  et~al.}{2012}]{Duffy2012}
{Duffy} A.~R.,  {Meyer} M.~J.,  {Staveley-Smith} L.,  {Bernyk} M.,  {Croton}
  D.~J.,  {Koribalski} B.~S.,  {Gerstmann} D.,   {Westerlund} S.,  2012,
  \mn@doi [\mnras] {10.1111/j.1365-2966.2012.21987.x}, \href
  {http://adsabs.harvard.edu/abs/2012MNRAS.426.3385D} {426, 3385}

\bibitem[\protect\citeauthoryear{{Einstein}}{{Einstein}}{1916}]{Einstein1916}
{Einstein} A.,  1916, \mn@doi [Annalen der Physik] {10.1002/andp.19163540702},
  \href {http://adsabs.harvard.edu/abs/1916AnP...354..769E} {354, 769}

\bibitem[\protect\citeauthoryear{{Elahi}, {Thacker}  \& {Widrow}}{{Elahi}
  et~al.}{2011}]{Elahi2011}
{Elahi} P.~J.,  {Thacker} R.~J.,   {Widrow} L.~M.,  2011, \mn@doi [\mnras]
  {10.1111/j.1365-2966.2011.19485.x}, \href
  {http://adsabs.harvard.edu/abs/2011MNRAS.418..320E} {418, 320}

\bibitem[\protect\citeauthoryear{{Elia}, {Ludlow}  \& {Porciani}}{{Elia}
  et~al.}{2012}]{Elia2012}
{Elia} A.,  {Ludlow} A.~D.,   {Porciani} C.,  2012, \mn@doi [\mnras]
  {10.1111/j.1365-2966.2012.20572.x}, \href
  {http://adsabs.harvard.edu/abs/2012MNRAS.421.3472E} {421, 3472}

\bibitem[\protect\citeauthoryear{{Erdo{\v g}du} et~al.,}{{Erdo{\v g}du}
  et~al.}{2006}]{Erdogdu2006}
{Erdo{\v g}du} P.,  et~al., 2006, \mn@doi [\mnras]
  {10.1111/j.1365-2966.2006.11049.x}, \href
  {http://adsabs.harvard.edu/abs/2006MNRAS.373...45E} {373, 45}

\bibitem[\protect\citeauthoryear{{Foreman-Mackey}, {Hogg}, {Lang}  \&
  {Goodman}}{{Foreman-Mackey} et~al.}{2013}]{ForemanMackey2013}
{Foreman-Mackey} D.,  {Hogg} D.~W.,  {Lang} D.,   {Goodman} J.,  2013, \mn@doi
  [\pasp] {10.1086/670067}, \href
  {http://adsabs.harvard.edu/abs/2013PASP..125..306F} {125, 306}

\bibitem[\protect\citeauthoryear{{Fry} \& {Gaztanaga}}{{Fry} \&
  {Gaztanaga}}{1993}]{Fry1993}
{Fry} J.~N.,  {Gaztanaga} E.,  1993, \mn@doi [\apj] {10.1086/173015}, \href
  {http://adsabs.harvard.edu/abs/1993ApJ...413..447F} {413, 447}

\bibitem[\protect\citeauthoryear{{Giovanelli} et~al.,}{{Giovanelli}
  et~al.}{2005}]{Giovanelli2005}
{Giovanelli} R.,  et~al., 2005, \mn@doi [\aj] {10.1086/497431}, \href
  {http://adsabs.harvard.edu/abs/2005AJ....130.2598G} {130, 2598}

\bibitem[\protect\citeauthoryear{{Gorski}, {Davis}, {Strauss}, {White}  \&
  {Yahil}}{{Gorski} et~al.}{1989}]{Gorski1989}
{Gorski} K.~M.,  {Davis} M.,  {Strauss} M.~A.,  {White} S.~D.~M.,   {Yahil} A.,
   1989, \mn@doi [\apj] {10.1086/167771}, \href
  {http://adsabs.harvard.edu/abs/1989ApJ...344....1G} {344, 1}

\bibitem[\protect\citeauthoryear{{Heitmann}, {Lawrence}, {Kwan}, {Habib}  \&
  {Higdon}}{{Heitmann} et~al.}{2014}]{Heitmann2014}
{Heitmann} K.,  {Lawrence} E.,  {Kwan} J.,  {Habib} S.,   {Higdon} D.,  2014,
  \mn@doi [\apj] {10.1088/0004-637X/780/1/111}, \href
  {http://adsabs.harvard.edu/abs/2014ApJ...780..111H} {780, 111}

\bibitem[\protect\citeauthoryear{{Heymans} et~al.,}{{Heymans}
  et~al.}{2012}]{Heymans2012}
{Heymans} C.,  et~al., 2012, \mn@doi [\mnras]
  {10.1111/j.1365-2966.2012.21952.x}, \href
  {http://adsabs.harvard.edu/abs/2012MNRAS.427..146H} {427, 146}

\bibitem[\protect\citeauthoryear{{Hildebrandt} et~al.,}{{Hildebrandt}
  et~al.}{2017}]{Hildebrandt2017}
{Hildebrandt} H.,  et~al., 2017, \mn@doi [\mnras] {10.1093/mnras/stw2805},
  \href {http://adsabs.harvard.edu/abs/2017MNRAS.465.1454H} {465, 1454}

\bibitem[\protect\citeauthoryear{{Hinshaw} et~al.,}{{Hinshaw}
  et~al.}{2013}]{Hinshaw2013}
{Hinshaw} G.,  et~al., 2013, \mn@doi [\apjs] {10.1088/0067-0049/208/2/19},
  \href {http://adsabs.harvard.edu/abs/2013ApJS..208...19H} {208, 19}

\bibitem[\protect\citeauthoryear{{Hong} et~al.,}{{Hong}
  et~al.}{2013}]{Hong2013}
{Hong} T.,  et~al., 2013, \mn@doi [\mnras] {10.1093/mnras/stt555}, \href
  {http://adsabs.harvard.edu/abs/2013MNRAS.432.1178H} {432, 1178}

\bibitem[\protect\citeauthoryear{{Hong} et~al.,}{{Hong}
  et~al.}{2014}]{Hong2014}
{Hong} T.,  et~al., 2014, \mn@doi [\mnras] {10.1093/mnras/stu1774}, \href
  {http://adsabs.harvard.edu/abs/2014MNRAS.445..402H} {445, 402}

\bibitem[\protect\citeauthoryear{{Howlett}, {Lewis}, {Hall}  \&
  {Challinor}}{{Howlett} et~al.}{2012}]{Howlett2012}
{Howlett} C.,  {Lewis} A.,  {Hall} A.,   {Challinor} A.,  2012, \mn@doi [\jcap]
  {10.1088/1475-7516/2012/04/027}, \href
  {http://adsabs.harvard.edu/abs/2012JCAP...04..027H} {4, 027}

\bibitem[\protect\citeauthoryear{{Howlett}, {Ross}, {Samushia}, {Percival}  \&
  {Manera}}{{Howlett} et~al.}{2015}]{Howlett2015a}
{Howlett} C.,  {Ross} A.~J.,  {Samushia} L.,  {Percival} W.~J.,   {Manera} M.,
  2015, \mn@doi [\mnras] {10.1093/mnras/stu2693}, \href
  {http://adsabs.harvard.edu/abs/2015MNRAS.449..848H} {449, 848}

\bibitem[\protect\citeauthoryear{{Howlett}, {Staveley-Smith}  \&
  {Blake}}{{Howlett} et~al.}{2017}]{Howlett2017}
{Howlett} C.,  {Staveley-Smith} L.,   {Blake} C.,  2017, \mn@doi [\mnras]
  {10.1093/mnras/stw2466}, \href
  {http://adsabs.harvard.edu/abs/2017MNRAS.464.2517H} {464, 2517}

\bibitem[\protect\citeauthoryear{{Huchra} et~al.,}{{Huchra}
  et~al.}{2012}]{Huchra2012}
{Huchra} J.~P.,  et~al., 2012, \mn@doi [\apjs] {10.1088/0067-0049/199/2/26},
  \href {http://adsabs.harvard.edu/abs/2012ApJS..199...26H} {199, 26}

\bibitem[\protect\citeauthoryear{{Hui} \& {Greene}}{{Hui} \&
  {Greene}}{2006}]{Hui2006}
{Hui} L.,  {Greene} P.~B.,  2006, \mn@doi [\prd] {10.1103/PhysRevD.73.123526},
  \href {http://adsabs.harvard.edu/abs/2006PhRvD..73l3526H} {73, 123526}

\bibitem[\protect\citeauthoryear{{Hunter}}{{Hunter}}{2007}]{Hunter2007}
{Hunter} J.~D.,  2007, \mn@doi [Computing in Science and Engineering]
  {10.1109/MCSE.2007.55}, \href
  {http://adsabs.harvard.edu/abs/2007CSE.....9...90H} {9, 90}

\bibitem[\protect\citeauthoryear{{Huterer}, {Shafer}  \& {Schmidt}}{{Huterer}
  et~al.}{2015}]{Huterer2015}
{Huterer} D.,  {Shafer} D.~L.,   {Schmidt} F.,  2015, \mn@doi [\jcap]
  {10.1088/1475-7516/2015/12/033}, \href
  {http://adsabs.harvard.edu/abs/2015JCAP...12..033H} {12, 033}

\bibitem[\protect\citeauthoryear{{Jaffe} \& {Kaiser}}{{Jaffe} \&
  {Kaiser}}{1995}]{Jaffe1995}
{Jaffe} A.~H.,  {Kaiser} N.,  1995, \mn@doi [\apj] {10.1086/176551}, \href
  {http://adsabs.harvard.edu/abs/1995ApJ...455...26J} {455, 26}

\bibitem[\protect\citeauthoryear{{Jarrett}, {Chester}, {Cutri}, {Schneider},
  {Skrutskie}  \& {Huchra}}{{Jarrett} et~al.}{2000}]{Jarrett2000}
{Jarrett} T.~H.,  {Chester} T.,  {Cutri} R.,  {Schneider} S.,  {Skrutskie} M.,
   {Huchra} J.~P.,  2000, \mn@doi [\aj] {10.1086/301330}, \href
  {http://adsabs.harvard.edu/abs/2000AJ....119.2498J} {119, 2498}

\bibitem[\protect\citeauthoryear{{Jennings}, {Baugh}  \& {Hatt}}{{Jennings}
  et~al.}{2015}]{Jennings2015}
{Jennings} E.,  {Baugh} C.~M.,   {Hatt} D.,  2015, \mn@doi [\mnras]
  {10.1093/mnras/stu2043}, \href
  {http://adsabs.harvard.edu/abs/2015MNRAS.446..793J} {446, 793}

\bibitem[\protect\citeauthoryear{{Johnson} et~al.,}{{Johnson}
  et~al.}{2014}]{Johnson2014}
{Johnson} A.,  et~al., 2014, \mn@doi [\mnras] {10.1093/mnras/stu1615}, \href
  {http://adsabs.harvard.edu/abs/2014MNRAS.444.3926J} {444, 3926}

\bibitem[\protect\citeauthoryear{{Johnston} et~al.,}{{Johnston}
  et~al.}{2008}]{Johnston2008}
{Johnston} S.,  et~al., 2008, \mn@doi [Experimental Astronomy]
  {10.1007/s10686-008-9124-7}, \href
  {http://adsabs.harvard.edu/abs/2008ExA....22..151J} {22, 151}

\bibitem[\protect\citeauthoryear{{Kaiser}}{{Kaiser}}{1987}]{Kaiser1987}
{Kaiser} N.,  1987, \mn@doi [\mnras] {10.1093/mnras/227.1.1}, \href
  {http://adsabs.harvard.edu/abs/1987MNRAS.227....1K} {227, 1}

\bibitem[\protect\citeauthoryear{{Kochanek} et~al.,}{{Kochanek}
  et~al.}{2001}]{Kochanek2001}
{Kochanek} C.~S.,  et~al., 2001, \mn@doi [\apj] {10.1086/322488}, \href
  {http://adsabs.harvard.edu/abs/2001ApJ...560..566K} {560, 566}

\bibitem[\protect\citeauthoryear{{Koda} et~al.,}{{Koda}
  et~al.}{2014}]{Koda2014}
{Koda} J.,  et~al., 2014, \mn@doi [\mnras] {10.1093/mnras/stu1610}, \href
  {http://adsabs.harvard.edu/abs/2014MNRAS.445.4267K} {445, 4267}

\bibitem[\protect\citeauthoryear{{Koribalski}}{{Koribalski}}{2012}]{Koribalski2012}
{Koribalski} B.~S.,  2012, \mn@doi [\pasa] {10.1071/AS12030}, \href
  {http://adsabs.harvard.edu/abs/2012PASA...29..359K} {29, 359}

\bibitem[\protect\citeauthoryear{{Lewis}, {Challinor}  \& {Lasenby}}{{Lewis}
  et~al.}{2000}]{Lewis2000}
{Lewis} A.,  {Challinor} A.,   {Lasenby} A.,  2000, \mn@doi [\apj]
  {10.1086/309179}, \href {http://adsabs.harvard.edu/abs/2000ApJ...538..473L}
  {538, 473}

\bibitem[\protect\citeauthoryear{{Linder} \& {Cahn}}{{Linder} \&
  {Cahn}}{2007}]{Linder2007}
{Linder} E.~V.,  {Cahn} R.~N.,  2007, \mn@doi [Astroparticle Physics]
  {10.1016/j.astropartphys.2007.09.003}, \href
  {http://adsabs.harvard.edu/abs/2007APh....28..481L} {28, 481}

\bibitem[\protect\citeauthoryear{{Ma}, {Gordon}  \& {Feldman}}{{Ma}
  et~al.}{2011}]{Ma2011}
{Ma} Y.-Z.,  {Gordon} C.,   {Feldman} H.~A.,  2011, \mn@doi [\prd]
  {10.1103/PhysRevD.83.103002}, \href
  {http://adsabs.harvard.edu/abs/2011PhRvD..83j3002M} {83, 103002}

\bibitem[\protect\citeauthoryear{{Macaulay}, {Feldman}, {Ferreira}, {Jaffe},
  {Agarwal}, {Hudson}  \& {Watkins}}{{Macaulay} et~al.}{2012}]{Macaulay2012}
{Macaulay} E.,  {Feldman} H.~A.,  {Ferreira} P.~G.,  {Jaffe} A.~H.,  {Agarwal}
  S.,  {Hudson} M.~J.,   {Watkins} R.,  2012, \mn@doi [\mnras]
  {10.1111/j.1365-2966.2012.21629.x}, \href
  {http://adsabs.harvard.edu/abs/2012MNRAS.425.1709M} {425, 1709}

\bibitem[\protect\citeauthoryear{{Malmquist}}{{Malmquist}}{1924}]{Malmquist1924}
{Malmquist} K.~G.,  1924, Meddelanden fran Lunds Astronomiska Observatorium
  Serie II, \href {http://adsabs.harvard.edu/abs/1924MeLuS..32....3M} {32, 3}

\bibitem[\protect\citeauthoryear{{Masters}, {Springob}, {Haynes}  \&
  {Giovanelli}}{{Masters} et~al.}{2006}]{Masters2006}
{Masters} K.~L.,  {Springob} C.~M.,  {Haynes} M.~P.,   {Giovanelli} R.,  2006,
  \mn@doi [\apj] {10.1086/508924}, \href
  {http://adsabs.harvard.edu/abs/2006ApJ...653..861M} {653, 861}

\bibitem[\protect\citeauthoryear{{Masters}, {Springob}  \& {Huchra}}{{Masters}
  et~al.}{2008}]{Masters2008}
{Masters} K.~L.,  {Springob} C.~M.,   {Huchra} J.~P.,  2008, \mn@doi [\aj]
  {10.1088/0004-6256/135/5/1738}, \href
  {http://adsabs.harvard.edu/abs/2008AJ....135.1738M} {135, 1738}

\bibitem[\protect\citeauthoryear{{Masters}, {Crook}, {Hong}, {Jarrett},
  {Koribalski}, {Macri}, {Springob}  \& {Staveley-Smith}}{{Masters}
  et~al.}{2014}]{Masters2014}
{Masters} K.~L.,  {Crook} A.,  {Hong} T.,  {Jarrett} T.~H.,  {Koribalski}
  B.~S.,  {Macri} L.,  {Springob} C.~M.,   {Staveley-Smith} L.,  2014, \mn@doi
  [\mnras] {10.1093/mnras/stu1225}, \href
  {http://adsabs.harvard.edu/abs/2014MNRAS.443.1044M} {443, 1044}

\bibitem[\protect\citeauthoryear{{Mueller}, {Percival}, {Linder}, {Alam},
  {Zhao}, {S{\'a}nchez}  \& {Beutler}}{{Mueller} et~al.}{2016}]{Mueller2016}
{Mueller} E.-M.,  {Percival} W.,  {Linder} E.,  {Alam} S.,  {Zhao} G.-B.,
  {S{\'a}nchez} A.~G.,   {Beutler} F.,  2016, preprint, \href
  {http://adsabs.harvard.edu/abs/2016arXiv161200812M} {} (\mn@eprint {arXiv}
  {1612.00812})

\bibitem[\protect\citeauthoryear{{Phillips}}{{Phillips}}{1993}]{Phillips1993}
{Phillips} M.~M.,  1993, \mn@doi [\apjl] {10.1086/186970}, \href
  {http://adsabs.harvard.edu/abs/1993ApJ...413L.105P} {413, L105}

\bibitem[\protect\citeauthoryear{{Planck Collaboration} et~al.,}{{Planck
  Collaboration} et~al.}{2016}]{Planck2016}
{Planck Collaboration} et~al., 2016, \mn@doi [\aap]
  {10.1051/0004-6361/201525830}, \href
  {http://adsabs.harvard.edu/abs/2016A%26A...594A..13P} {594, A13}

\bibitem[\protect\citeauthoryear{{Poole} et~al.,}{{Poole}
  et~al.}{2015}]{Poole2015}
{Poole} G.~B.,  et~al., 2015, \mn@doi [\mnras] {10.1093/mnras/stv314}, \href
  {http://adsabs.harvard.edu/abs/2015MNRAS.449.1454P} {449, 1454}

\bibitem[\protect\citeauthoryear{{Press} \& {Schechter}}{{Press} \&
  {Schechter}}{1974}]{Press1974}
{Press} W.~H.,  {Schechter} P.,  1974, \mn@doi [\apj] {10.1086/152650}, \href
  {http://adsabs.harvard.edu/abs/1974ApJ...187..425P} {187, 425}

\bibitem[\protect\citeauthoryear{{Riess} et~al.,}{{Riess}
  et~al.}{2016}]{Riess2016}
{Riess} A.~G.,  et~al., 2016, \mn@doi [\apj] {10.3847/0004-637X/826/1/56},
  \href {http://adsabs.harvard.edu/abs/2016ApJ...826...56R} {826, 56}

\bibitem[\protect\citeauthoryear{{Scrimgeour} et~al.,}{{Scrimgeour}
  et~al.}{2016}]{Scrimgeour2016}
{Scrimgeour} M.~I.,  et~al., 2016, \mn@doi [\mnras] {10.1093/mnras/stv2146},
  \href {http://adsabs.harvard.edu/abs/2016MNRAS.455..386S} {455, 386}

\bibitem[\protect\citeauthoryear{{Silberman}, {Dekel}, {Eldar}  \&
  {Zehavi}}{{Silberman} et~al.}{2001}]{Silberman2001}
{Silberman} L.,  {Dekel} A.,  {Eldar} A.,   {Zehavi} I.,  2001, \mn@doi [\apj]
  {10.1086/321663}, \href {http://adsabs.harvard.edu/abs/2001ApJ...557..102S}
  {557, 102}

\bibitem[\protect\citeauthoryear{{Song} \& {Percival}}{{Song} \&
  {Percival}}{2009}]{Song2009}
{Song} Y.-S.,  {Percival} W.~J.,  2009, \mn@doi [\jcap]
  {10.1088/1475-7516/2009/10/004}, \href
  {http://adsabs.harvard.edu/abs/2009JCAP...10..004S} {10, 004}

\bibitem[\protect\citeauthoryear{{Springel}}{{Springel}}{2005}]{Springel2005}
{Springel} V.,  2005, \mn@doi [\mnras] {10.1111/j.1365-2966.2005.09655.x},
  \href {http://adsabs.harvard.edu/abs/2005MNRAS.364.1105S} {364, 1105}

\bibitem[\protect\citeauthoryear{{Springob}, {Haynes}, {Giovanelli}  \&
  {Kent}}{{Springob} et~al.}{2005}]{Springob2005}
{Springob} C.~M.,  {Haynes} M.~P.,  {Giovanelli} R.,   {Kent} B.~R.,  2005,
  \mn@doi [\apjs] {10.1086/431550}, \href
  {http://adsabs.harvard.edu/abs/2005ApJS..160..149S} {160, 149}

\bibitem[\protect\citeauthoryear{{Springob}, {Masters}, {Haynes}, {Giovanelli}
  \& {Marinoni}}{{Springob} et~al.}{2007}]{Springob2007}
{Springob} C.~M.,  {Masters} K.~L.,  {Haynes} M.~P.,  {Giovanelli} R.,
  {Marinoni} C.,  2007, \mn@doi [\apjs] {10.1086/519527}, \href
  {http://adsabs.harvard.edu/abs/2007ApJS..172..599S} {172, 599}

\bibitem[\protect\citeauthoryear{{Springob} et~al.,}{{Springob}
  et~al.}{2014}]{Springob2014}
{Springob} C.~M.,  et~al., 2014, \mn@doi [\mnras] {10.1093/mnras/stu1743},
  \href {http://adsabs.harvard.edu/abs/2014MNRAS.445.2677S} {445, 2677}

\bibitem[\protect\citeauthoryear{{Springob} et~al.,}{{Springob}
  et~al.}{2016}]{Springob2016}
{Springob} C.~M.,  et~al., 2016, \mn@doi [\mnras] {10.1093/mnras/stv2648},
  \href {http://adsabs.harvard.edu/abs/2016MNRAS.456.1886S} {456, 1886}

\bibitem[\protect\citeauthoryear{{Tinker}, {Weinberg}  \& {Zheng}}{{Tinker}
  et~al.}{2006}]{Tinker2006}
{Tinker} J.~L.,  {Weinberg} D.~H.,   {Zheng} Z.,  2006, \mn@doi [\mnras]
  {10.1111/j.1365-2966.2006.10114.x}, \href
  {http://adsabs.harvard.edu/abs/2006MNRAS.368...85T} {368, 85}

\bibitem[\protect\citeauthoryear{{Tully} \& {Fisher}}{{Tully} \&
  {Fisher}}{1977}]{Tully1977}
{Tully} R.~B.,  {Fisher} J.~R.,  1977, \aap, \href
  {http://adsabs.harvard.edu/abs/1977A%26A....54..661T} {54, 661}

\bibitem[\protect\citeauthoryear{{Tully}, {Courtois}  \& {Sorce}}{{Tully}
  et~al.}{2016}]{Tully2016}
{Tully} R.~B.,  {Courtois} H.~M.,   {Sorce} J.~G.,  2016, \mn@doi [\aj]
  {10.3847/0004-6256/152/2/50}, \href
  {http://adsabs.harvard.edu/abs/2016AJ....152...50T} {152, 50}

\bibitem[\protect\citeauthoryear{{Watkins} \& {Feldman}}{{Watkins} \&
  {Feldman}}{2015}]{Watkins2015}
{Watkins} R.,  {Feldman} H.~A.,  2015, \mn@doi [\mnras] {10.1093/mnras/stv651},
  \href {http://adsabs.harvard.edu/abs/2015MNRAS.450.1868W} {450, 1868}

\bibitem[\protect\citeauthoryear{{Wojtak}, {Davis}  \& {Wiis}}{{Wojtak}
  et~al.}{2015}]{Wojtak2015}
{Wojtak} R.,  {Davis} T.~M.,   {Wiis} J.,  2015, \mn@doi [\jcap]
  {10.1088/1475-7516/2015/07/025}, \href
  {http://adsabs.harvard.edu/abs/2015JCAP...07..025W} {7, 025}

\bibitem[\protect\citeauthoryear{{da Cunha} et~al.,}{{da Cunha}
  et~al.}{2017}]{daCunha2017}
{da Cunha} E.,  et~al., 2017, preprint, \href
  {http://adsabs.harvard.edu/abs/2017arXiv170601246D} {} (\mn@eprint {arXiv}
  {1706.01246})

\bibitem[\protect\citeauthoryear{{de la Torre} et~al.,}{{de la Torre}
  et~al.}{2013}]{delaTorre2013}
{de la Torre} S.,  et~al., 2013, \mn@doi [\aap] {10.1051/0004-6361/201321463},
  \href {http://adsabs.harvard.edu/abs/2013A%26A...557A..54D} {557, A54}

\makeatother
\end{thebibliography}
\bibliographystyle{mnras}

\appendix

\end{document}